%% file: main.tex
\documentclass[camera]{jpaper}
\usepackage[nocompress]{cite}

\usepackage{soul}

\usepackage{amsmath}
\usepackage{graphicx}
\usepackage{titlesec}

\titlespacing*{\section}{0pt}{3pt}{-1pt}
\titlespacing*{\subsection}{0pt}{3pt}{1pt}
\titlespacing*{\subsubsection}{0pt}{1pt}{0pt}

\usepackage{array}
\makeatletter
\let\MYcaption\@makecaption
\makeatother

\usepackage[font=footnotesize]{subcaption}

\makeatletter
\let\@makecaption\MYcaption
\makeatother

\usepackage{fixltx2e}

\usepackage{dblfloatfix}
\usepackage[nolessnomore, italic]{mathastext}
\usepackage[T1]{fontenc}
\usepackage[usenames,dvipsnames,svgnames,table]{xcolor}
\usepackage{multirow}
\usepackage{hhline}
\usepackage[normalem]{ulem}
\usepackage{setspace}
\usepackage{indentfirst}
\usepackage{footmisc}

\usepackage{pifont}

\usepackage{fancyhdr}
\usepackage[normalem]{ulem}
\PassOptionsToPackage{hyphens}{url}
\usepackage{hyperref}
\usepackage{url}
\usepackage{footnote}
\makesavenoteenv{tabular}
\makesavenoteenv{table}
\usepackage{booktabs}

\usepackage{amsmath,amssymb,amsfonts}
\usepackage{graphicx}
\usepackage{textcomp}
\usepackage{xcolor}
\usepackage{fancyhdr}
 \usepackage{amsmath}
 \usepackage{footnote}
 \usepackage{algorithm}
\usepackage{algpseudocode}
 \usepackage{multicol}
 \usepackage{siunitx}
 \usepackage{xspace}

\newcommand{\ignore}[1]{}
\usepackage{fancyhdr}
\usepackage{color}
\usepackage{soul}
\usepackage{multirow}
\usepackage{enumitem}
\usepackage{tikz}
\usepackage[redeflists]{IEEEtrantools}
\usepackage{makecell}

\definecolor{darkgreen}{rgb}{0.0, 0.2, 0.13}

\newcommand\lois[1]{\noindent{\color{black} #1}}

\newcommand{\sr}[1]{{\color{black}#1}}

\newcommand{\ja}[1]{{\color{black} {#1}}}
\newcommand{\jb}[1]{{\color{black} {#1}}}
\newcommand{\jk}[1]{{\color{black}#1}}

\newcommand{\jkz}[1]{{\color{black}#1}}

\newcommand{\jc}[1]{{\color{black} {#1}}}
\newcommand{\lon}[1]{{\color{black} {#1}}}

\newcommand{\jkx}[1]{{\color{black}#1}}
\newcommand{\lo}[1]{{\color{black} {#1}}}
\newcommand{\jd}[1]{{\color{black} {#1}}}

\newcommand{\je}[1]{{\color{black} {#1}}}
\newcommand{\jkc}[1]{{\color{black}#1}}
\newcommand{\jkv}[1]{{\color{black}#1}}
\newcommand{\jkb}[1]{{\color{black}#1}}

\newcommand{\hjred}[1]{{\color{black} {#1}}}
\newcommand{\hjorg}[1]{{\color{black} {#1}}}

\newcommand{\hjg}[1]{{\color{black} {#1}}}
\newcommand{\hjp}[1]{{\color{black} {#1}}}
\newcommand{\hjb}[1]{{\color{black} {#1}}}
\newcommand{\hjov}[1]{{\color{black} {#1}}}

\newcommand{\hja}[1]{{\color{black} {#1}}}
\newcommand{\hjz}[1]{{\color{black} {#1}}}

\newcommand{\hly}[2][yellow]{{%
    {#2}
    }%
}
\newcommand{\hlp}[2][pink]{{%
    {#2}}%
}
\newcommand{\hlo}[2][orange]{{%
    {#2}}%
}

\newcommand{\hlb}[2][cyan]{{%
    {#2}}%
}

\newcommand{\tech}{FlexWatts\xspace}
\newcommand{\modl}{PDNspot\xspace}

\newif\ifcameraready
\camerareadyfalse

\ifcameraready

\else

\fi

\newcommand{\affilIntel}[0]{\textsuperscript{$\star$}}
\newcommand{\affilNTU}[0]{\textsuperscript{$\ddagger$}}
\newcommand{\affilETH}[0]{\textsuperscript{\S}}
\newcommand{\affilTECH}[0]{\textsuperscript{$\dagger$}}

\definecolor{amber}{rgb}{1.0, 0.49, 0.0}
\definecolor{darkbyzantium}{rgb}{0.36, 0.22, 0.33}
\definecolor{darkseagreen}{rgb}{0.56, 0.74, 0.56}
\definecolor{darkspringgreen}{rgb}{0.09, 0.45, 0.27}
\definecolor{dollarbill}{rgb}{0.52, 0.73, 0.4}

\ifcameraready
\else
 \fancypagestyle{firststyle}
 {
   \fancyhead[C]{\textcolor{MidnightBlue}{\emph{Version \versionnum~---~\today, \ampmtime}}}
   \fancyfoot[C]{\thepage}
 }
\fi

\newcommand{\versionnum}[0]{6.0 \today~@ 6:20 AM CET}

\usepackage{caption}
\begin{document}
\bstctlcite{IEEEexample:BSTcontrol} 


\title{\vspace{-28pt}\ja{{\tech}:} A Power- and Workload-\ja{A}ware Hybrid Power Delivery Network for  Energy-\ja{E}fficient \ja{Microprocessors}}


%


\author{
{Jawad Haj-Yahya\affilETH}\qquad~~~%
{Mohammed Alser\affilETH}\qquad~~~%
{Jeremie \jkc{S.} Kim\affilETH}\qquad~~~ 
\vspace{2pt}
{Lois Orosa\affilETH}\\
{Efraim Rotem\affilIntel}\qquad%
{Avi Mendelson\affilTECH\affilNTU}\qquad%
{Anupam Chattopadhyay\affilNTU}\qquad%
\vspace{6pt}
{Onur Mutlu\affilETH}\\%
\emph{{\affilETH ETH Z{\"u}rich \qquad   \affilIntel Intel  \qquad \affilTECH Technion%
\qquad \affilNTU Nanyang Technological University}}%
\vspace{-5pt}%
}



%


\maketitle
\thispagestyle{plain} 
\pagestyle{plain}

\setstretch{0.93}
\renewcommand{\footnotelayout}{\setstretch{0.9}}

\input{body}

\section*{Acknowledgments} We thank the anonymous reviewers of \ja{MICRO 2020} for feedback and the SAFARI group members for
feedback and the stimulating intellectual environment they provide.



%


\SetTracking
 [ no ligatures = {f},
 outer kerning = {*,*} ]
 { encoding = * }
 { -40 } 

{

  \let\OLDthebibliography\thebibliography
  \renewcommand\thebibliography[1]{
    \OLDthebibliography{#1}
    \setlength{\parskip}{3pt}
    \setlength{\itemsep}{2pt}
  }
  \bibliographystyle{IEEEtranS}
  \bibliography{references}
}

\end{document}

%% file: body.tex
\begin{abstract}
\vspace*{-.5em}
\jk{Modern client processors typically use one of three commonly-used power delivery network (PDN) architectures: 1) \emph{motherboard voltage regulators} (MBVR), 2) \emph{integrated voltage regulators} (IVR), and 3) \emph{low dropout voltage regulators} (LDO). We observe that the energy-efficiency of \jc{each of} these PDNs varies with the processor power (e.g., thermal design power (TDP) \ja{and dynamic} power-state) and workload characteristics (e.g.,  workload type and computational intensity). This leads to energy-inefficienc\jc{y} and  performance loss, as modern client processors operate across a wide spectrum of power consumption and execute a wide variety of workloads.}

\jk{To address this inefficiency, we} propose \emph{\tech}, a hybrid adaptive PDN for modern client processors whose goal is to provide high energy-efficiency across the processor's wide \ja{range} of power \ja{consumption}  and workloads. \jk{{\tech} provides high energy-efficiency by intelligently and dynamically allocating PDNs to \jc{processor} domains depending on the processor's power consumption and workload. {\tech} is based on} \emph{three} key ideas.
First, {\tech} combines IVRs and LDOs in a novel way \jk{to share multiple \jkc{on-chip and off-chip} resources and \jd{thus reduce cost, as well as board and die area overheads.} This hybrid PDN is allocated} for processor domains with a \emph{wide} power consumption range (e.g., CPU cores and graphics engines) \jk{and  \jc{it} \emph{dynamically} switches between two modes: \texttt{IVR-Mode} and \texttt{LDO-Mode}\jc{,} depending on the power consumption.}
Second, for \jk{all other processor domains (that have a \jb{low and} narrow power range, e.g., \jc{the} IO domain),} {\tech} statically allocates \jb{off-chip VRs\jd{,} which have high energy-efficiency for \jb{low and} narrow power ranges}.
Third, {\tech} introduces a novel prediction algorithm that automatically switches the hybrid PDN to the mode (\texttt{IVR-Mode} or \texttt{LDO-Mode}) \ja{that is the} most beneficial based on processor power \jk{consumption} and workload characteristics. 

\ja{To evaluate the \jd{tradeoffs of} PDNs, we develop \ja{and open-source} \emph{\modl}, the first validated \jd{architectural} PDN model that enables quantitative analysis of PDN metrics. 
\jk{Using {\modl}, we} evaluate {\tech} \jc{on} a wide variety of SPEC CPU2006, graphics (3DMark06), and battery life \jd{(e.g., video playback)} workloads  \jk{against} IVR, the state-of-the-art \ja{PDN} in modern client processors. 
\jc{For} a $4W$ \jk{thermal design power (TDP)} processor, {\tech} improves the average performance of \jc{the} SPEC CPU2006 and 3DMark06 workloads by $22\%$ and $25\%$, respectively. 
For battery life workloads, {\tech} reduces the average power consumption of video playback by $11\%$ across all tested TDPs (4W--50W). {\tech} has comparable cost and area overhead to IVR. 
\ja{We conclude that {\tech} \jk{provides} high energy-efficiency across \jc{a modern client} processor's wide \ja{range} of power \ja{consumption} and \jc{wide variety of} workloads\jc{,} with minimal overhead. } }

\end{abstract}


\section{Introduction}
Architecting an efficient \emph{power delivery network} (PDN) for client processors (e.g.,  tablets, laptops, desktops) is a well-known challenge that has been hotly debated in industry and academia in recent years.
\jk{Due to multiple constraints, a modern client processor typically \ja{implements} only \emph{one} of} three \ja{types of} commonly-used PDNs: 
1) \lowercase{\emph{Motherboard Voltage Regulators}} (MBVR \cite{9_rotem2011power,jahagirdar2012power,11_fayneh20164,haj2019comprehensive}), 2) \lowercase{\emph{Low Dropout Voltage Regulators}} (LDO  \cite{singh20173,singh2018zen,burd2019zeppelin,beck2018zeppelin,toprak20145,sinkar2013low}), and 3) \lowercase{\emph{Integrated Voltage Regulators}} (IVR  \cite{2_burton2014fivr,5_nalamalpu2015broadwell,tam2018skylake,icelake2020}).
We find that the energy-efficiency of each \ja{of the three different} commonly-used PDN \ja{types} varies \ja{differently} with the processor power (e.g., thermal design power (TDP\footnote{As the processor dissipates power, the temperature of the silicon junction ($T_j$) increases, ($T_j$) should be kept below the maximum junction temperature ($T_j{_{max}}$). Overheating may cause permanent damage to the processor. \ja{Hence, every processor has a thermal design power (TDP) limit}.}) \jd{and dynamic} power-state) and workload characteristics (e.g., workload type and computational intensity).
Particularly, each PDN is designed for energy-\ja{efficient operation at} a different TDP, power-state, workload type, and  workload computational intensity. 
\jk{This leads to energy-inefficienc\jc{y} and performance loss as modern client processors operate across a \emph{wide} range of power consumption and execute a wide variety of workloads}. 

Architects of modern  client processors typically build a \emph{single} PDN architecture (i.e., MBVR, IVR, or LDO) that supports \emph{all} TDPs of  \ja{a} client processor \ja{family}  for two reasons. 
First,  \ja{doing so} allows \ja{system} manufacturers to configure a processor's TDP (known as configurable TDP \cite{cTDP,cTDP2,jahagirdar2012power} or cTDP) to enable \jd{the}  processor to operate at higher or lower performance levels, depending on the available cooling capacity and desired power consumption. For example, the Intel Skylake processor uses an MBVR PDN \cite{tam2018skylake,21_doweck2017inside} for all TDP ranges (from  $3W$ \cite{intel_skl__3_5} to $91W$ \cite{intel_skl_91}) \jk{and recent AMD  client processors use an LDO PDN \cite{singh20173,singh2018zen,burd2019zeppelin,beck2018zeppelin,amd_zen2_10,amd_zen2_54}}, \ja{while} enabling cTDP \cite{intel_skl__3_5,intel_skl_91}.
Second, \ja{it} reduces non-recurring engineering (NRE \cite{magarshack2003system}) cost and \jd{design} complexity to  allow competitive product prices \jd{and \jkc{enable meeting of} strict time-to-market requirements}.

Modern  client processors operate \jk{across} a \emph{wide power range} (i.e., the range \ja{of power consumption} between  \ja{under} light-load and heavy-load) for two reasons. 
First, modern workloads have a wide range of \ja{computational} intensity (\ja{leading to} between tens of milliwatts \jc{of power consumption}, e.g., \ja{for an idle workload that is in} Connected-Standby \ja{power-mode} \cite{haj_connected_standby}, to tens of watts on average, e.g., \ja{for} a workload that activates Turbo Boost \cite{rotem2015intel}). 
Second, processors must support multiple market segments that have \ja{very} different TDPs. For example, the recent \jd{Intel} Skylake processor architecture can scale from near\jc{ly} $3W$ \cite{intel_skl__3_5} of TDP (for passively\jd{-}cooled small systems, e.g., \ja{a} tablet) up to $91W$ \cite{intel_skl_91} of TDP (\ja{for a} high-performance desktop computer). The recent AMD client processors follow similar trends \cite{singh20173,singh2018zen,burd2019zeppelin,beck2018zeppelin,amd_zen2_10,amd_zen2_54}.


\ja{Based on} our \ja{empirical} evaluations, 
we \ja{find} that a single PDN architecture, which supports a wide power range is energy-inefficient. 
For instance, the IVR PDN is energy-inefficient \ja{for} low\ja{-}TDP processors (e.g., tablets, \ja{convertible laptop-tablet\jc{s}}), while the MBVR and the LDO PDNs are energy-inefficient for high\ja{-}TDP processors (e.g., high performance laptops, desktops). 
We also observe that \emph{even} if we build a dedicated PDN \ja{matching the} TDP of  \ja{the} processor, e.g., IVR PDN for high-TDP processors and MBVR or LDO PDN \ja{for} low-TDP processors, these processors will still suffer from significant energy inefficiency because 
1) the IVR PDN is energy-inefficient in high-TDP processors when running a \ja{computationally} light \ja{work}load, 
2) a low-TDP processor can \ja{potentially} execute \ja{computationally} heavy \ja{workloads} that exceed the TDP, e.g., \ja{via} Turbo Boost \cite{rotem2015intel},
and 3) the TDP of modern  client processors can be dynamically configured using cTDP \cite{cTDP,cTDP2}. 

\ja{Various works} focus on improving \jk{the} processor PDN using various techniques (e.g., thermal-aware voltage regulators (VRs) \cite{khatamifard2017thermogater},
re-configurable PDN \cite{arch1}, 
VR phase scaling \cite{asghari2016vr}, 
VR efficiency-aware power management \cite{bai2017voltage}, 
on-chip VRs for fast DVFS \cite{kim2008system,arch3,yan2012agileregulator},
voltage stacking \cite{arch2,pal2019architecting,gopireddy2019designing}, 
PDNs for waferscale processors \cite{pal2019architecting},
voltage noise reduction \cite{reddi2010voltage,arch4,thomas2016core,shevgoor2013quantifying,grochowski2002microarchitectural,gupta2008decor,haj2015compiler,leng2015gpu,reddi2009voltage,miller2012vrsync}, voltage noise modeling \cite{zhang2014architecture,zou2017ivory},
multiple voltage domains \cite{rotem2009multiple,yan2010leveraging}, 
voltage optimizations \cite{arch7}, and adaptive DVFS \cite{arch8,ccakmak2015cyclic}). 
These works focus on \ja{\jc{adapting} power management} techniques \ja{\jd{that} already exist } in modern \ja{client} processors \ja{(}such as voltage noise reduction and modeling, power management techniques that optimize VR efficiency, using fast VRs for better DVFS, utilizing on-chip VRs for building multiple voltage  domains to improve energy-efficiency\ja{)}\jc{,} \ja{but} they 
do not \ja{alleviate} the \ja{inherent} energy inefficiencies of commonly-used PDNs
in \emph{client} processors \ja{due to operating across a \emph{wide} range of power and \jd{wide variety of} workloads.}

In this paper, we propose  \emph{\tech}, a power- and workload-aware hybrid adaptive PDN whose \textbf{goal} is to maintain high energy efficiency in a modern  client processor \ja{throughout} the processor's wide spectrum of power and workloads with a low bill of materials (BOM\footnote{Given a specific product, a BOM is a list of its immediate components \ja{with} which it is built and the \jd{components'} relationships.}\cite{jiao2000generic}) and board area overhead. {\tech} is based on \emph{three} \textbf{key ideas}.
First, {\tech} combines IVRs and LDOs in a novel way \jk{to share multiple \jkc{on-chip and off-chip} resources and} \jd{thus} reduce BOM, \jd{as well as}  board and die area \ja{overheads}. This hybrid PDN \jk{is allocated for processor domains with a wide power consumption range (e.g., CPU cores and graphics engines) and \jd{it} dynamically switches between two modes, \texttt{IVR-Mode} and \texttt{LDO-Mode}, depending on the power consumption. For} example, when a domain operates \ja{under} high power conditions (e.g., high TDP, power-hungry applications), it uses \ja{the PDN in } \texttt{IVR-Mode}. Otherwise (e.g., low TDP, light-load), \ja{it uses the PDN in} \texttt{LDO-Mode}. Second, \jc{for} \jk{all other processor domains (that have a \jb{low and} narrow power range, e.g., \ja{the} IO domain)}, {\tech} statically allocates \jb{off-chip VRs \jc{that} have high energy-efficiency for low and narrow power ranges.} 
Third, {\tech} introduces a new prediction algorithm that automatically switches the hybrid PDN to the mode (i.e., \texttt{IVR-Mode} or \texttt{LDO-Mode}) that \jc{is predicted to be the} most beneficial based on processor power \ja{consumption} and workload characteristics\ja{.}

\begin{sloppypar}
To assess the tradeoffs of commonly-used PDNs, and architect a PDN that is highly efficient in the metrics of interest (e.g., energy consumption, performance, board area, BOM), an accurate \emph{architecture-level} quantitative analysis of these metrics is needed. 
Unfortunately, no model \ja{or tool} is available to the computer architecture research community \ja{for such analysis}. To this end, we develop \emph{\modl}, a validated \jd{architectural} \ja{open-source} PDN \ja{framework} whose goal is to enable architects to study the tradeoffs of various PDN\jc{s}. {\modl} provides a versatile \ja{framework} that enables multi-dimensional architecture-space exploration of modern processor PDNs. {\modl} \ja{evaluates} the effect of multiple PDN parameters, TDP, and workloads on the metrics of interest. \ja{We open-source {\modl}\jc{\cite{pdnspot_source}}.}
\end{sloppypar}

\jd{Using {\modl}, we evaluate {\tech} on a wide variety of SPEC CPU2006, graphics (3DMark06), and battery life \jd{(e.g., video playback)} workloads  against IVR~\cite{2_burton2014fivr}, the state-of-the-art PDN in modern client processors. 
For a $4W$ TDP processor, {\tech} improves the average performance of the SPEC CPU2006 and 3DMark06 workloads by $22\%$ and $25\%$, respectively. 
For battery life workloads, {\tech} reduces the average power consumption of video playback by $11\%$ across all tested TDPs ($4W$--$50W$). {\tech} has comparable BOM and area overhead to IVR.}

This paper makes the following major \textbf{contributions}:
\begin{itemize}

\item We introduce {\tech}, a novel adaptive hybrid PDN that maintains high efficiency \jd{and high performance} in metrics of interest \jk{in} client processors across \ja{a wide} spectrum of power \ja{consumption} and workloads. To our knowledge, {\tech} is the \emph{first} hybrid PDN to use two types of \emph{on-chip} voltage regulators (IVR and LDO) \jk{to simultaneously leverage the advantages of both}.

\item  We develop a versatile framework, {\modl}, that enables multi-dimensional architecture-level exploration of modern processor PDNs. To our knowledge, {\modl} is the first \ja{tool that can evaluate}  the effects of multiple PDN parameters, TDP, and workloads characteristics on prominent system metrics such as  energy consumption, performance, board area, and bill of materials (BOM). 
\ja{We open-source {\modl}\jc{\cite{pdnspot_source}}.}


\item We provide a thorough experimental evaluation of the power, performance, area, and BOM of IVR, MBVR, LDO, and {\tech} PDNs across various processor TDPs and workloads. \ja{Our} evaluation shows that our \jc{new} adaptive hybrid PDN, {\tech}, \ja{provides} \jc{large benefits} in metrics of interest \jc{(performance, energy, cost, area)} with  minimal overhead\ja{, compared to the state-of-the-art PDN.}


\end{itemize}

\section{Background}\label{sec:background}
We \ja{provide} the necessary background on \jk{the architecture of a modern client processor and its power delivery network (PDN), the electrical system that provides supply voltage to the transistors within an integrated circuit via voltage regulators. We also explain some of the parameters (e.g., tolerance band and load-line) that affect the} system\ja{-}level efficiency of PDNs.

\begin{figure*}[b]
  \begin{center}
  \includegraphics[trim=0.6cm 0.6cm 0.6cm .6cm,clip=true, width=0.9\linewidth]{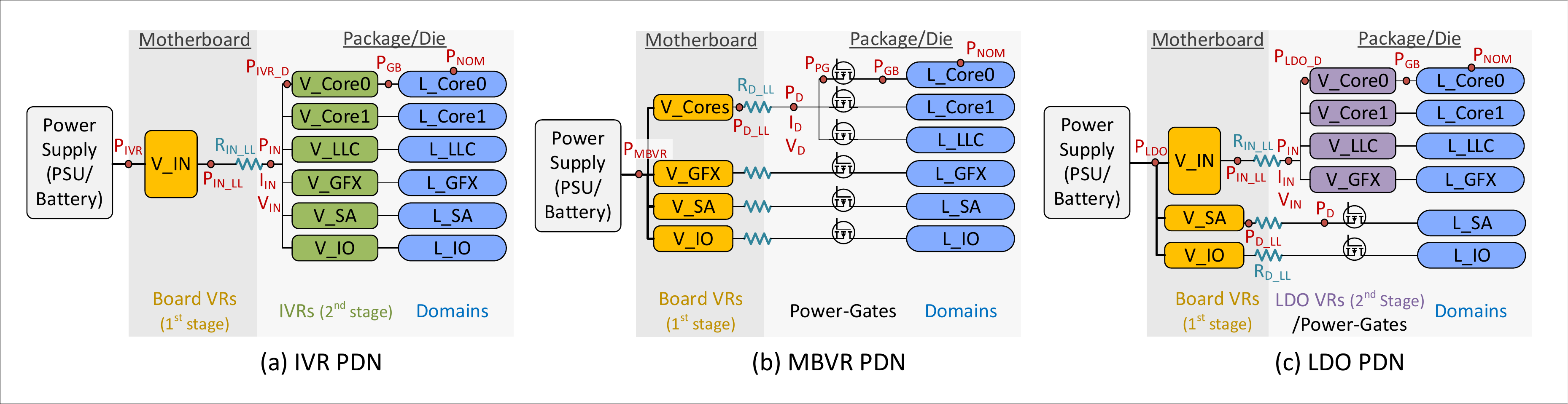}\\
  \caption{The three commonly-used PDNs in client processors. The processor consists of six loads: two \ja{CPU} cores, a last-level-cache (LLC), graphics engines (GFX),  system-agent (SA), \ja{and IO}. 
  (a) The IVR PDN uses one off-chip VR (V\_IN) and six different on-chip IVRs \ja{(V\_Core0/1, V\_LLC, V\_GFX, V\_SA and V\_IO)}. 
  (b) The \textbf{MBVR} PDN uses four off-chip VRs \ja{(V\_Cores, V\_GFX, V\_SA and V\_IO)} and \ja{six} on-chip power-gates. 
  (c) The \textbf{LDO} PDN uses three off-chip VRs (V\_IN, V\_SA and V\_IO), \jkx{four} on-chip LDO \jc{VRs} \ja{(V\_Core0/1, V\_LLC, V\_GFX)}, and \jc{two} on-chip  power-gates.  
  }\label{fig3_three_schems}
  \end{center} 
 \vspace*{-5mm}
\end{figure*}

\subsection{PDNs in Modern Client Processors}
\noindent \textbf{Architecture.} To illustrate the usage of a PDN in modern client processors, we first summarize the architecture of Intel's client processor~\cite{2_burton2014fivr, rotem2012power, anati2016inside, burres2015intel, meinerzhagen2018energy} in Table \ref{tbl:sys_arch}. Similar architectures are widely used for modern processors from various vendors, such as AMD\jc{, IBM,} and ARM \cite{singh20173,singh2018zen,burd2019zeppelin,beck2018zeppelin,toprak20145,qcomm2018,nikolskiy2016efficiency}.

\begin{table}[!ht]
\centering
\caption{Summary of the processor architecture }
\label{tbl:sys_arch}
\resizebox{\linewidth}{!}{%
\begin{tabular}{llll}
\hline
\textbf{Domain}                                                            & \textbf{Description}                                                                                                                                                                                  \\  \toprule
\begin{tabular}[c]{@{}l@{}}Two CPU Cores\\ (Core 0/1)\end{tabular}    & \begin{tabular}[c]{@{}l@{}}\jc{Single clock domain to all cores. Clock} frequency can\\ scale from 0.8GHz to 4GHz\end{tabular}                               \\ \hline
\begin{tabular}[c]{@{}l@{}}Graphics Engines\\  (GFX)\end{tabular}  & \begin{tabular}[c]{@{}l@{}}GFX frequency can scale from 0.1GHz to 1.2GHz\end{tabular}                                                               \\ \hline
\begin{tabular}[c]{@{}l@{}}Last Level Cache\\  (LLC)\end{tabular} & \begin{tabular}[c]{@{}l@{}}The LLC size scales proportionally to the\\ \jc{CPU} core and graphics \jc{engine} frequencies\end{tabular}                                                                      \\ \hline
\begin{tabular}[c]{@{}l@{}}System-Agent\footnote{\jc{The System-Agent houses the traditional North Bridge and contains several modules such as the memory and IO \jkx{controllers~\cite{Skylake_die, usui2016dash,haj2020sysscale}}}.} \\ (SA)\end{tabular}      & \begin{tabular}[c]{@{}l@{}}The SA includes a memory controller, display controller, \\ IO fabric, and other IPs (e.g., Camera, PCIe,  Voice), each of\\ which operate at a fixed frequency (not scal\jc{ed} with load)\end{tabular} \\ \hline
\begin{tabular}[c]{@{}l@{}}Input/Output \\ (IOs)\end{tabular}     & \begin{tabular}[c]{@{}l@{}}Includes the processor IOs,  such as DDRIO, display\\ IO, \jc{which operate at} fixed frequencies\end{tabular}                                                             \\ \hline
\end{tabular}
}
\end{table}

\noindent \textbf{Power Delivery Networks.} 
\jk{The Power Delivery Network (PDN) is the electrical system that provides supply voltage to the transistors within an integrated circuit (IC) or domain (e.g., CPU core, graphics engine) in a processor.
The objective of a PDN in a processor is to provide a stable desired voltage to each processor domain\jc{.}} \ja{Particularly, a PDN should support three distinct capabilities:
1) supply a stable voltage to each processor domain, 2) provide transient current required by a processor domain, and 3) filter out the noise currents injected by a processor domain\ja{\cite{swaminathan2007power,15_jakushokas2010power,vaisband2013Hybrid}.}} 

\jk{A PDN consists of 
1) a \emph{power supply} (e.g., power supply unit (PSU) or battery), which provides high voltage (e.g., $7.2$--$20V$) to the motherboard,
2) \emph{voltage regulators} (VRs) (also known as DC–DC converters), used in either one or two stages to reduce the voltage level from the power supply to the desired operational voltage for a domain (typically $0.5$--$1.1V$), 
3) a \emph{network of interconnections}, which distributes the voltage from the voltage regulators to the PDN components and \jkc{processor domains},
4) \emph{decoupling capacitors} distributed on the motherboard, package, and die, which act as reservoirs to store charge and reduce voltage noise from instantaneous current draw, and
5) \emph{power-gates} to turn off a processor domain when it is idle.}
\jk{Before discussing the common PDN designs in more detail, we first discuss types of voltage regulators, an essential component in PDNs for converting voltage.}

\subsection{\ja{Voltage Regulators (VRs)}} \label{Buck_Converters}

\ja{The main objective of a voltage regulator (VR) is to convert the input voltage level to another voltage level. There are multiple types of VRs and each has pros and cons with respect to power conversion efficiency, voltage noise, \jb{design complexity and size}. In this section, we describe the switching VR (SVR), and the low dropout VR (LDO \jc{VR}), each of which are key  components \jb{(\jd{on-chip} and/or off-chip)} in modern client processor PDNs.} 

\noindent \textbf{Switching Voltage Regulator (SVR).}
Modern processors typically use a step-down SVR (i.e., \jk{a buck} converter\ja{\cite{22_hazucha2005233,perreault2009opportunities,kim2008system}}), which converts the input voltage level \jk{to a lower voltage level.} An SVR consists of an inductor, diode, capacitor, \jk{switch, and control modules}. Traditionally, SVRs are placed on the \emph{motherboard}. However, recent PDN designs \emph{integrate} SVRs into the chip package and die~\cite{2_burton2014fivr,5_nalamalpu2015broadwell,tam2018skylake,icelake2020}. 
The main advantage of an SVR over other types of VRs is its ability to maintain a high power conversion efficiency (typically ${>}80\%$) even if the output voltage is very different from the input voltage. 
Unfortunately, \jc{SVR} has four main disadvantages compared to other VR types: 1) complicated design, 2) high cost, 3) high voltage noise, and 4) \jc{it} requires \jk{a large difference in the} input/output voltage levels \cite{kazimierczuk2015pulse} (\jc{i.e., voltage headroom,} e.g., a minimum difference of $0.6V$ for an input voltage of $1.8V$). 

\noindent \textbf{Low Dropout Voltage Regulator (LDO VR).}
\ja{An LDO VR is a type of linear voltage regulator \cite{milliken2007full,15_jakushokas2010power,luria2016dual} that consists of a power switch, a differential amplifier (error amplifier), and resistors. 
The LDO VR has four advantages over an SVR: an LDO VR 1) is immune to switching noise due to the absence of capacitors, 2) has a simpler and smaller design as it does not include large inductors, 3) can regulate the output voltage even when the input voltage \ja{level} is very close to the output voltage \ja{level}, \jc{4)} even operate in \textit{bypass-mode} \cite{singh2018zen}, in which the input voltage signal is \jc{directly} connected to the output to avoid voltage regulation, and \jc{5}) can have higher efficiency than an SVR when the input voltage level is very close to the output voltage level (e.g., input/output voltage of $1V$/$0.9V$).  However, the main disadvantage of the LDO VR is its inefficiency in converting the input voltage if it is very different from the output voltage (e.g., input/output voltage of $1V$/$0.5V$).} 



\subsection{\ja{Power Delivery Network} } \label{pdn_bg}
\jk{
Fig.~\ref{fig3_three_schems} shows \jkx{the high-level organization of} each of the three commonly-used PDNs in modern client \jkx{processors:} 1) \lowercase{\emph{Integrated Voltage Regulator}} (IVR \cite{2_burton2014fivr,5_nalamalpu2015broadwell,tam2018skylake,icelake2020}; Fig.~\ref{fig3_three_schems}(a)),
2) \lowercase{\emph{Motherboard Voltage Regulator}} (MBVR \cite{9_rotem2011power,jahagirdar2012power,11_fayneh20164,haj2019comprehensive}; Fig.~\ref{fig3_three_schems}(b)), 
and 3) \lowercase{\emph{Low Dropout Voltage Regulator}} (LDO VR \cite{singh20173,singh2018zen,burd2019zeppelin,beck2018zeppelin,toprak20145,sinkar2013low}; Fig.~\ref{fig3_three_schems}(c)).}

\noindent \textbf{Integrated Voltage Regulator (IVR) PDN.}
\jk{The IVR PDN is a \jc{state-of-the-art} PDN in modern client processors and is used \jc{in} Intel's~4th, 5th, and 10th  generation Core processors~\cite{2_burton2014fivr, 5_nalamalpu2015broadwell,icelake2020}.
The IVR PDN integrates most of the SVR components (i.e., diodes, capacitors, control modules, and switches) into the processor die while some components are placed on the package (\jd{e.g.,} interconnections) and off-chip (\jkc{e.g., inductors}).}
\jk{Since circuit elements in modern processors cannot tolerate the high input voltage of a power supply ($7.2$--$20V$) due to \jc{their} small process technology \jc{node} size, the IVR PDN regulates voltage in \emph{two-stages}, as illustrated in Fig. \ref{fig3_three_schems}(a). 
The first stage of voltage conversion \ja{is handled by a single motherboard SVR}  (i.e., $V\_IN$ VR), which converts input voltage from the power supply unit (PSU) or battery ($7.2$--$20V$) to a level typically less than $2V$ (e.g., $1.8V$). The second stage is handled by an integrated SVR (i.e., IVR), which is a sequential buck converter that converts the input voltage (i.e., output of the first stage VR) to the desired voltage level (typically $0.5$--$1.1V$) of a processor domain (e.g., a CPU core). In a processor, multiple IVRs are used (e.g., six as shown in Fig. \ref{fig3_three_schems}(a)) to supply different voltage levels to each processor domain.

The IVR PDN has two main advantages \jc{over other PDNs:} 1) it enables fast voltage level changes, 2) it reduces a chip's input (i.e., output of the first stage VR \jc{into} the processor die) current by using a high input voltage level (e.g., $1.8V$ compared to $0.5$--$1.1V$ using a traditional MBVR), thereby reducing $I^2R$ power losses, and reduces the maximum current (i.e., $Icc_{max}$) requirement of the first stage VR.
However, the IVR PDN has three main disadvantages \jc{over other PDNs:} 1) low power-conversion efficiency in computationally light workloads due to the two-stage voltage regulation \cite{haj2019comprehensive}, 2) high design complexity as it is normally designed along with the chip\jc{,} which adds \jd{extra} design constraints and consumes silicon die area  \cite{naffziger2016integrated}, and 3) higher sensitivity to \jkx{\emph{di/dt}} noise than the MBVR PDN due to a limited amount of decoupling capacitors available on the processor's die \cite{naffziger2016integrated}.}

\noindent \textbf{Motherboard Voltage Regulator (MBVR) PDN.} 
\jk{The MBVR PDN is the traditional PDN for processors and is used in} Intel's~2nd, 3rd, 6th, 7th, 8th and 9th  generation Core processors \cite{9_rotem2011power,jahagirdar2012power,11_fayneh20164,kabylake_2019,coffeelake_2020}.  As shown in Fig.~\ref{fig3_three_schems}(b), \jk{the} MBVR PDN uses several \jk{one-stage} motherboard \jd{SVRs} and multiple on-chip power-gates.  
\jk{An MBVR PDN has four advantages over other PDNs: 1) it decouples the VR design from the processor design, \ja{thereby reducing system design complexity}, 2) \jk{heat generated} due to \jd{VR} power conversion losses is kept outside the processor chip, 3) it enables placing enough decoupling capacitors on motherboard, package and die \jd{(due to the long path from processor die to the off-chip VR)} to reduce voltage noise, and 4) it is efficient \jc{at} executing computationally light workloads.} However, the MBVR PDN has two major disadvantages: 1) voltage level changes are slow as the VR is far from the load (i.e., processor \jd{domain}), and 2) \jc{\jkx{computationally-intensive} (high current) workloads suffer} high $I^2R$ power losses  due to high processor input current and high impedance (load-line) on the path from the board VRs to the processor domains.


\noindent \textbf{Low Dropout Voltage Regulator (LDO) PDN.}
\jk{The LDO PDN is used in AMD's recent Zen \cite{singh20173,singh2018zen,beck2018zeppelin} processors. 
As shown in Fig. \ref{fig3_three_schems}(c), the LDO PDN statically allocates two types of VRs to different domains based on their power demands: it allocates 1) one-stage motherboard \jd{SVRs} (similar to MBVR PDN) to domains with a \jb{low and} narrow power range (e.g., IO and SA) and 2) two-stage VRs for domains with wide power \jc{range} (e.g., CPU cores, graphics engines, and LLC). The first stage is a single motherboard SVR (i.e., $V\_IN$ VR) and the second stage is an integrated LDO VR. Multiple LDO VRs are used (e.g., four as shown in Fig. \ref{fig3_three_schems}(c)) which supply different voltage levels to each of the processor domains.
For the two-stage VR, the processor's power management unit adjusts $V\_IN$ to the maximum voltage required across all domains. For domains that require the same voltage level as the input voltage, the domain's LDO VR operates in \emph{bypass-mode} to avoid voltage regulation by simply connecting the input voltage signal to the output. For other domains that require a lower voltage, the LDO VR adjusts the input voltage by operating in \emph{regulation-mode}. For idle domains, the LDO VR acts as a \emph{power-gate}.

The LDO \jc{PDN} has three advantages over other PDNs: \jc{it} 1) requires less board area compared to the MBVR PDN, 2) is simpler than the IVR PDN as the integration of an LDO VR into the die is simpler than \jc{that of} an SVR, 3) has higher  power-conversion efficiency than an IVR PDN when running computationally light workloads.
However, the LDO PDN has two main disadvantage\jc{s} compared to other PDNs: 1) low power-conversion efficiency in computationally intensive workloads due to the high processor input current and high impedance (load-line) on the path from the board VRs to the processor domains, and 2) higher design complexity than MBVR as it is designed along with the chip\jc{,} which adds \jkx{extra} design constraints and complexity to the power management algorithms.}


\subsection{VR and PDN Parameters}\label{sec:vr_params}
\noindent \textbf{Power-Conversion Efficiency ($\eta$).}
The ratio of the total output power ($P_{out}$) of a VR to the total input power ($P_{in}$) is known as \emph{Efficiency} ($\eta$) as given in Equation \ref{efficiency}.

\begin{equation}\label{efficiency}
\resizebox{.55\hsize}{!}{$
	\textnormal{Efficiency}=\eta=\frac{P_{out}}{P_{in}} =\frac{P_{out}}{P_{out}+ P_{loss}} 
$}
\end{equation}

For an SVR, power-conversion efficiency is not constant, but rather a function of: 1) the load current and 2) the input and output voltages \cite{haj2018power,haj2018energy,gough2015cpu,bai2017voltage}. 
The LDO \jc{VR} power-conversion efficiency, $\eta_{LDO}$\jd{,} is the ratio of the desired output voltage, $V_{out}$, to the input voltage, $V_{in}$, times the LDO \jc{VR}  \emph{current efficiency} (typically  around 99\% in a modern LDO \jc{VR}~\cite{luria2016dual,huang2016fully}), thus $\eta_{LDO}\approx V_{out}/V_{in}$.

The power-conversion efficiency is also defined for the \emph{entire PDN}, also known as the PDN \textbf{end-to-end power-conversion efficiency (ETEE)}. ETEE of a PDN \jd{at a given time} is the ratio between the total \jc{load's} \je{nominal} \jkc{power} (i.e., the sum of all \jc{loads'} \je{nominal}  power\footnote{\jd{\jkc{A load's} nominal power at a given time is a function of \jkc{the} load's 1) power state (e.g., active vs. idle), 2) activity factor, 3) frequency, 4) nominal voltage, and 5) temperature \cite{haj2016fine,haj2018energy,haj2018power,gough2015cpu}.}}) and the effective power consumed by the main power supply (e.g., battery, PSU).

\noindent \textbf{\hlb{VR Tolerance Band (TOB).}}
The \jkb{tolerance band (TOB)} of a VR~\cite{1_regulator200611} is \hlb{the maximum voltage variation for the VR across temperature}, manufacturing variation, and age factors (e.g., $V_{TOB} =25mV$). The standard \jkb{VR TOB} can be sliced into three main categories: controller tolerance, current sense variation, and voltage ripple. The supply voltage is maintained at a higher value than the nominal voltage required by the load\jc{,} to compensate for TOB voltage variations. This excess voltage \jc{due to the TOB} leads to wasted power that cannot be utilized by the load.

\noindent \textbf{Application Ratio (AR).} AR is a term used in power/performance modeling  to quantify the \emph{computational intensity} of a workload \cite{gough2015cpu}. AR describes the switching rate of a processor component (e.g., CPU core, graphics engine, IO) \jc{for} a workload when compared to the highest possible power, $P_{peak}$, that can be consumed by the most computational\jc{ly-}intensive workload (i.e., \ja{also known as} \jc{the} \emph{power-virus} workload \cite{5_nalamalpu2015broadwell,li2009mcpat,ganesan2010system}). 
AR and $P_{peak}$ can be estimated 1) offline using power modeling tools such as McPAT\cite{li2009mcpat}, SYMPO \cite{ganesan2010system}  or Intel's Blizzard~\cite{anshumali2010circuit}), \hlp{and 2) at runtime using \emph{activity sensors} implemented in the processor components \mbox{\cite{ananthakrishnan2019controlling,vogman2018method,ardanaz2018hierarchical,shrall2017controlling,rotem2016system,linda2014dynamic,burns2016method,fetzer2015managing}}}.

\noindent \textbf{\hly{Load-line.}}\label{sec:load_line}
\hly{The load-line or adaptive voltage positioning~\cite{14_module2009and} is a model that describes the voltage and current relationship under a given system impedance ($R_{LL}$). This relationship is defined as: $V_{cc} =  V_{IN} - V_{TOB} - R_{LL}\cdot I_{cc}$~where $V_{cc}$ and $I_{cc}$ are the voltage and current at the load, respectively.} $V_{TOB}$ \hlb{is the tolerance band (TOB) voltage variation} and $V_{IN}$ \jc{is} the input voltage to the system. From this equation, we can see that the voltage at the load input ($V_{cc}$) decreases when the current of the load ($I_{cc}$) increases \jkx{(e.g., \jkc{when} running a workload with a high AR). Therefore, to keep the voltage at the load ($V_{cc}$) above a minimum functional voltage under even the most computationally-intensive workload (i.e., \emph{power-virus}\je{~\cite{5_nalamalpu2015broadwell,li2009mcpat,ganesan2010system}}, for which AR=1), the input voltage ($V_{IN}$) is set to a level that provides enough guardband.} 

\section{{\modl} }\label{sec:model}
We develop {\modl}, a framework that models the three commonly-used PDNs in modern  client processors\jkx{, evaluating} multiple metrics of interest \jkz{(i.e., \jc{performance, energy, BOM, and board area})}. {\modl} provides a versatile \ja{framework} that enables multi-dimensional architecture-space exploration of modern processor PDNs. {\modl} \ja{evaluates} the effect of multiple PDN parameters, TDP, and workloads on the metrics of interest.
In this section, we present \jk{the core models of {\modl}}:
1) an end-to-end power-conversion efficiency (ETEE) \ja{model for each PDN} that we use to assess the average power and current consumption of \ja{a} PDN,
2) board area and BOM models, 
and 3) a performance model of the processor that we use to assess \jkz{each PDN's impact on performance.}

\subsection{End-to-End Power-Conversion  Efficiency
\\(ETEE) Modeling}\label{sec:models}

We present three high-level \emph{power models}. Each model takes multiple inputs (main inputs tabulated in Table \ref{tab:params}) to calculate the end-to-end power consumption of a domain \ja{(shown on the right side of each PDN in Fig.~\ref{fig3_three_schems}),} starting from nominal power of each load ($P_{NOM}$,  in Fig.~\ref{fig3_three_schems}) until the power supply \ja{(shown on the left side of each PDN in Fig.~\ref{fig3_three_schems})}. The calculations follow the symbols shown in Fig.~\ref{fig3_three_schems} on each PDN to estimate the total power (i.e., $P_{IVR}$, $P_{MBVR}$, and $P_{LDO}$)  consumed by the main power supply (i.e., PSU or battery).

We calculate the end-to-end power-conversion efficiency (ETEE) of each PDN as the ratio of the total input power of the PDN (i.e., the sum of the nominal \jkz{input} power \ja{of all loads}, $\sum P_{NOM}$) to the total effective power (i.e., $P_{IVR}$, $P_{MBVR}$, and $P_{LDO}$) consumed by the main power supply. \jk{We begin by discussing MBVR PDN modeling as it is the simplest PDN.}

\begin{table}[h]
\centering
\caption{Main parameters used in our {\modl} models}
\label{tab:params}
\resizebox{\linewidth}{!}{%
\begin{tabular}{|l||c|c|c|}
\hline
\textbf{Parameter}         & \textbf{IVR}     & \textbf{MBVR}                    & \textbf{LDO}               \\ \Xhline{3\arrayrulewidth}
Load-line Impedance $R_{LL}$ (m\si{\ohm})      & $IN=1$           & $Cores,GFX, SA, IO=2.5,2.5,7,4$  & $IN,SA,IO=1.25,7,4$        \\ \hline
VR Tolerance Band $TOB$ (mV)                 & $18$--$22$       & $18$--$20$                       & $16$--$18$                 \\ \hline
On-chip VR Efficiency $\eta$ $(\%)$        & $81\%$--$88\%$   & ---                              & $(V_{out}/V_{in})\cdot99.1\%$  \\ \hline
Off-chip VR Efficiency $\eta$ $(\%)$       & \multicolumn{3}{c|}{$\eta_{IN,GFX,SA,IO}(V_{in},V_{out},I_{out},$power-state$)=72\%$--$93\%$} \\ \hline
Leakage Fraction $F_L$ $(\%)$      & \multicolumn{3}{c|}{$20\%$--$45\%$ \jk{depending} on the domain}                        \\ \hline
Cores Nom. Power $P_{NOM}$ (W)            & \multicolumn{3}{c|}{$0.6W$--$30W$ for TDP range $4$--$50W$}                      \\ \hline
LLC Nom. Power $P_{NOM}$ (W)              & \multicolumn{3}{c|}{$0.5W-4W$ for  TDP range $4$--$50W$}                         \\ \hline
GFX Nom. Power $P_{NOM}$ (W)         & \multicolumn{3}{c|}{$0.58W$--$29.4W$ for TDP range $4$--$50W$}                   \\ \hline
PG Impedance $R_{PG}$ (m\si{\ohm}) & \multicolumn{3}{c|}{$1$--$2\jkz{m\si{\ohm}}$ \jk{depending} on the domain}                              \\ \hline
\end{tabular}%
}
\end{table}

\noindent \textbf{MBVR PDN Power Modeling.}
In order to calculate the total power consumption of the MBVR, denoted by ($P_{MBVR}$), we first calculate $P_{GB}$\ja{, which} is the power \jk{consumption after applying a} voltage guardband  on the nominal power $P_{NOM}$. This voltage guardband, $V_{GB}$, guarantees proper circuit timing across \hlb{voltage variations (\mbox{\text{$V_{TOB}$}} explained in Sec.} \ref{sec:vr_params}). The \textit{leakage} and \textit{dynamic} power consumption scale differently as voltage increases \ja{from $V_{NOM}$ to $V_{NOM}+V_{GB}$ (i.e., when nominal voltage, $V_{NOM}$, is increased by a voltage guardband, $V_{GB}$)}. The dynamic power consumption is proportional to the voltage squared (i.e., $(\frac{V_{NOM}+V_{GB}}{V_{NOM}})^2$), while the leakage power consumption scales exponentially with voltage and depends on several other parameters such as threshold voltage, \hlb{temperature}, and other design and fabrication characteristics~\cite{haj2018power,haj2018energy,15_jakushokas2010power,gough2015cpu,haj2016fine}. As an approximation, we use a model based on polynomial curve fitting, where leakage power scales polynomially with supply voltage \ja{(i.e., $(\frac{V_{NOM}+V_{GB}}{V_{NOM}})^\delta$)}. We validate our model with measurements on a commercial \ja{client} processor (Intel Core~i7-6600U Processor \ja{\cite{intel6600U}}). Assuming the same \hlb{temperature}, the leakage power scales by the power of ${\delta}=\sim$2.8 proportional to voltage scaling. We assume a leakage fraction ($F_L$) of $45\%$ for the graphics domain and $22\%$ for the rest (e.g., cores, LLC, SA) similarly to Rusu \textit{et al.}~\cite{4_rusu20145}. Therefore, $P_{GB}$ can be calculated with Equation \ref{eqn:PGB}.

\vspace*{-4mm}
\begin{equation}\label{eqn:PGB}
\resizebox{.9\hsize}{!}{$
	P_{GB}=P_{NOM}\cdot 
    \Big[ 
    F_L\cdot (\frac{V_{NOM}+V_{GB}}{V_{NOM}})^\delta+ 
    (1-F_L)\cdot (\frac{V_{NOM}+V_{GB}}{V_{NOM}})^2 
    \Big] 
$}
\end{equation}
\vspace*{-3mm}


For domains with power-gates (e.g., $L\_Core0/1$ and $L\_LLC$ in Fig. \ref{fig3_three_schems}(b)), there is an additional voltage drop on the power-gate ($V_{PG}$, e.g., $10mV$) due to its impedance ($R_{PG}$). The power consumption ($P_{PG}$ in Fig. \ref{fig3_three_schems}(b)), due to this increase in the \ja{supply} voltage, \ja{is calculated \jk{similarly} to Equation~\ref{eqn:PGB} (i.e., by assigning in the equation: $V_{PG}$, $P_{GB}$, ($V_{NOM}+V_{GB}$) instead of $V_{GB}$, $P_{NOM}$, $V_{NOM}$, respectively)}. 

\begin{sloppypar}

Next, we need to compensate \ja{for} the voltage drop on the load-line impedance ($R_{LL}$, \ja{discussed in Sec. \ref{sec:vr_params}}) by raising the on-board VR output voltage (i.e., \jk{applying} a voltage guardband). The voltage guardband needs to account for the maximum possible voltage drop, which is attained when the processor consumes the maximum possible power, $P_{peak}$, \jb{by \jkz{running} the most \jkx{computationally-intensive workload, which is} also known as a  \emph{power-virus} workload}~\cite{5_nalamalpu2015broadwell,li2009mcpat,ganesan2010system}. 
\ja{Next we attain,} ${P_{D}}$, the total power consumption of a group of domains which share the same off-chip VR (e.g., $\{Core0,Core1,LLC\}$, $\{GFX\}$), \ja{by summing all $P_{PG}$ \jk{values from each} domain,}. \ja{We use the} application ratio (AR, \ja{discussed in Sec. \ref{sec:vr_params}}), \ja{to} obtain $P_{peak}$ by scaling ${P_{D}}$ using the AR, i.e.,\hly{ ${P_{peak}}={P_{D}}/AR$}. The corresponding calculation for the voltage and power after \jk{accounting for the voltage drop on the} load-line impedance of each group of domains \ja{(i.e., $R_{D\_LL}$ in Fig. \ref{fig3_three_schems}(b))} is shown by Equations~\ref{eqn:VDLL} and~\ref{eqn:PDLL}, respectively.
\end{sloppypar}

\vspace*{-7mm}
\begin{multicols}{2}
\begin{equation}\label{eqn:VDLL}
\resizebox{.63\hsize}{!}{{$
    V_{D\_LL}= V_{D} + \frac{P_{peak}}{V_{D}}\cdot R_{D\_LL}
$}}
\end{equation}\break
\begin{equation}\label{eqn:PDLL}
\resizebox{.79\hsize}{!}{$
    P_{D\_LL}= V_{D\_LL}\cdot I_{D} = V_{D\_LL}\cdot \frac{P_{D}}{V_{D}}
$}
\end{equation}
\end{multicols}
\vspace*{-2mm}

The total power, $P_{MBVR}$, consumed from the battery/PSU is obtained by summing the effective power of each domain, which can be calculated by dividing the output power of each on-board VR by its power conversion efficiency ($\eta_D$) as shown in Equation \ref{eqn:PMBVR}.

\vspace*{-4mm}
\begin{equation}\label{eqn:PMBVR}
\resizebox{.25\hsize}{!}{$
	P_{MBVR}=\sum \frac{P_{D\_LL}}{\eta_D}
$}
 \end{equation}

\noindent \textbf{IVR PDN Power Modeling.}
Using the same approach \jk{for modeling MBVR PDN power consumption, we calculate the total power of an IVR PDN,} $P_{IVR}$, consumed from the battery/PSU\jkz{,} \ja{as shown in Fig. \ref{fig3_three_schems}(a)}. We calculate $P_{GB}$ by applying a voltage guardband due to \hlb{the \ja{VR tolerance band (i.e., TOB, discussed in Sec. \ref{sec:vr_params})} using} Equation~\ref{eqn:PGB}. $P_{IVR\_D}$ (in Fig. \ref{fig3_three_schems}(a))  is the power \jk{consumption after accounting for} the IVR loss at a specific domain. Given the IVR \ja{power conversion} efficiency $\eta_{IVR}$, $P_{IVR\_D}$ can be calculated using Equation \ref{eqn:PIVRD}.

\vspace*{-1mm}
\begin{equation}\label{eqn:PIVRD}
\resizebox{.2\hsize}{!}{$
    P_{IVR\_D}= \frac{P_{GB}}{\eta_{IVR}}
$}
\end{equation}

\ja{Next we calculate $P_{IN}$ (shown in Fig. \ref{fig3_three_schems}(a)) by summing the power consumed by all domains connected to  $V\_{IN}$ VR (i.e., $P_{IN}=\sum P_{IVR\_D}$).}
Similarly to the MBVR PDN, the voltage \ja{($V_{IN\_LL}$)} and power \jk{consumption \ja{($P_{IN\_LL}$)} after accounting for} the \ja{voltage drop on the} load-line \ja{impedance (i.e., $R_{IN\_LL}$)} \jkz{are} calculated with Equations~\ref{eqn:VINLL} and \ref{eqn:PINLL}, respectively, \ja{whereas} ${{P_{IN_{peak}}}}={P_{IN}}/AR$.
\ja{Finally, we obtain the} total power ($P_{IVR}$) consumed from the battery/PSU by dividing the output power \ja{(i.e., $P_{IN\_LL}$)} of the $V_{IN}$ VR by the \ja{power conversion} efficiency of the $V_{IN}$ VR (\ja{i.e.,} $\eta_{IN}$)\jkz{,} as shown in Equation \ref{eqn:PIVR}.

\vspace*{-1mm}
\begin{equation}\label{eqn:VINLL}
\resizebox{.45\hsize}{!}{$
    V_{IN\_LL}= V_{IN} + \frac{{P_{IN_{peak}}}}{V_{IN}}\cdot R_{IN\_LL}
$}
\end{equation}

\vspace*{-8mm}
\begin{multicols}{2}
\begin{equation}\label{eqn:PINLL}
\resizebox{.72\hsize}{!}{$
    P_{IN\_LL} = V_{IN\_LL}\cdot \frac{P_{IN}}{V_{IN}}
$}
\end{equation}\break
\begin{equation}\label{eqn:PIVR}
\resizebox{.45\hsize}{!}{$
    P_{IVR}= \frac{P_{IN\_LL}}{\eta_{IN}}
$}
\end{equation}
\end{multicols}
\vspace*{-2mm}

\noindent \textbf{LDO PDN Power Modeling.}
Similarly to the other two \jkz{models, $P_{GB}$} (shown in Fig. \ref{fig3_three_schems}(c)) is calculated using Equation~\ref{eqn:PGB}. \ja{For the four domains with LDO \jkz{VRs} (i.e., $L\_Core0/1$, $L\_LLC$ and $L\_GFX$ domains), we calculate the power of each domain after including the LDO \jkz{VR} power conversion losses, denoted by $P_{LDO\_D}$ in Fig. \ref{fig3_three_schems}(c). $P_{LDO\_D}$ is obtained by dividing the output power of the LDO ($P_{GB}$) by the power conversion efficiency of the LDO ($\eta_{LDO}$) as shown in Equation~\ref{eqn:PLDO_D}. $\eta_{LDO}$ is the ratio of the desired output voltage to the input voltage \jk{multiplied by} the LDO \jkz{VR} current efficiency ($I_{effi}$, e.g., 99\%), as shown in Equation~\ref{eqn:LDO_EFF}. Next, we obtain the power that each LDO domain consumes from the shared VR  ($V\_{IN}$) using two steps. First, we sum the power of each LDO domain to obtain $P_{IN}$ (i.e.,  $P_{IN}=\sum P_{LDO\_D}$). Second, we calculate the power \jk{consumption  ($P_{IN\_LL}$) after accounting for} the \ja{voltage drop on the} load-line \ja{impedance (i.e., $R_{IN\_LL}$)} using Equations~\ref{eqn:VINLL} and \ref{eqn:PINLL} (similar to the calculations in IVR PDN \jk{power modeling}). } 


\vspace*{-7mm}
\begin{multicols}{2}
\begin{equation}\label{eqn:LDO_EFF}
\resizebox{.53\hsize}{!}{$
   \eta_{LDO} =\frac{V_{OUT}}{V_{IN}}\cdot I_{effi}
$}
\end{equation}\break
\begin{equation}\label{eqn:PLDO_D}
\resizebox{.46\hsize}{!}{$
    P_{LDO\_D}= \frac{P_{GB}}{\eta_{LDO}}
$}
\end{equation}
\end{multicols}
\vspace*{-1mm}

\ja{For domains that use motherboard VRs (i.e., $L\_SA$ and $L\_IO$), we calculate the power ($P_{D\_LL}$) that each of these domains consumes from the motherboard VRs (i.e., $V\_SA$ and $V\_IO$) using Equations~\ref{eqn:VDLL} and~\ref{eqn:PDLL} (similar to our calculations in MBVR PDN \jk{power modeling}).}
\ja{Finally, the total power (i.e., $P_{LDO}$) \jk{that} the LDO PDN consumes from the battery/PSU is calculated by summing the power that each motherboard VR consumes from the battery/PSU as shown in Equation~\ref{eqn:PLDO}.}

\vspace*{-1mm}
\begin{equation}\label{eqn:PLDO}
\resizebox{.35\hsize}{!}{$
	P_{LDO}= \frac{P_{IN\_LL}}{\eta_{IN}} +  \sum \frac{P_{D\_LL}}{\eta_D} 
$}
 \end{equation}

\subsection{Board Area and BOM Modeling} \label{sec:board_area}

The board area \ja{and BOM} of an off-chip VR \jk{are functions} of \ja{mainly} the maximum current ($Icc_{max}$) that the VR can support.
\ja{$Icc_{max}$ is the maximum current \jk{that the} VR must be electrically designed to support. Exceeding the the $Icc_{max}$ limit can result in irreversible damage to the VR or the processor's chip \cite{gough2015cpu,haj2018power,haj2018energy,skylakex,14_module2009and,naffziger2016integrated,wright2006characterization,zhang2014architecture,ma2014maximum}.}
A higher $Icc_{max}$ implies a larger VR and higher cost. 
VR sharing between multiple domains (e.g., the LDO PDN shares $V\_IN$ VR for cores, LLC, and graphics as shown in Fig. \ref{fig3_three_schems}(c)) effectively reduces the maximum current required, $Icc_{max}$, thereby reducing the area and BOM of the off-chip VR. 

To reduce system area and cost, many platforms use a \textit{power management integrated circuit} (PMIC \cite{pmic,huang2016ccm,shi2007highly}) that incorporates multiple VRs (and other functions) into one integrated circuit. In our model, the VR area and cost are calculated based on the $Icc_{max}$ requirements for each domain of \jk{a} PDN. We assume an optimized solution with a PMIC for \ja{processors with} TDPs up to $18W$ for all PDNs. Higher\ja{-}TDP \ja{processors} typically use a traditional voltage regulator module (VRM \cite{14_module2009and}) instead of a PMIC due to the high current requirements \ja{of these processors} \cite{huang2016ccm,shi2007highly}. We obtain the actual mapping between the $Icc_{max}$ and the area/cost directly from \emph{Texas Instruments} \jc{VR vendor}~\cite{TI_VRs}. 

\subsection{Processor Performance Modeling} \label{sec:perf_model}

\begin{sloppypar}
To understand the impact of PDN end-to-end power-conversion efficiency (ETEE) on workload \emph{performance} of a  client processor, we build a performance model. Our performance model aims to estimate the performance improvement of a CPU- (graphics-) intensive workload when increasing the power-budget allocated to the CPU cores (graphics engines).   
\end{sloppypar}

\begin{sloppypar}
We build the performance model of the compute domain (i.e., CPU cores and graphics engines) using  empirical measurements on a real system in three steps.  
First, we run a CPU- (graphics-) intensive \hjred{workload with high performance-scalability}\footnote{\hjred{We define performance scalability of a workload with respect to CPU frequency as the performance improvement the workload experiences with unit increase in frequency, as described in \cite{yasin2017performance,haj2015doee}. Modern processors predict the performance-scalability of a workload at runtime using performance counters \cite{yasin2017performance}. The performance-scalability metric is used by current power management algorithms, such as Intel's SpeedShift \cite{rotem2015intel} and EARtH \cite{efraim2012energy}, which first appeared in the Intel Skylake processor\jc{\cite{anati2016inside}}.}}, e.g., 416.gamess of SPEC \jc{CPU2006}~\cite{18_SPEC} (3DMark06\cite{17_3DMARK}), on a real \ja{Intel} Skylake system\ja{, whose} specifications \ja{are} in Table \ref{tbl:sys_setup}.
Second, we sweep the frequency of CPU cores (graphics engines) in steps of 100MHz (50MHz), the finest CPU core (graphics engine) frequency granularity that the Skylake architecture supports.
Third, we measure the total power consumption of the processor and log the increase in power consumption compared to the measurement done in the previous (i.e, lower) frequency. By doing so, we build power-frequency curves that we use along with the \hjred{workload's performance-scalability} to estimate performance as a function of power. 
\end{sloppypar}

\begin{figure}[h]
   \begin{center}
   \includegraphics[trim=0.8cm 0.75cm .9cm .8cm, clip=true,width=1.0\linewidth]{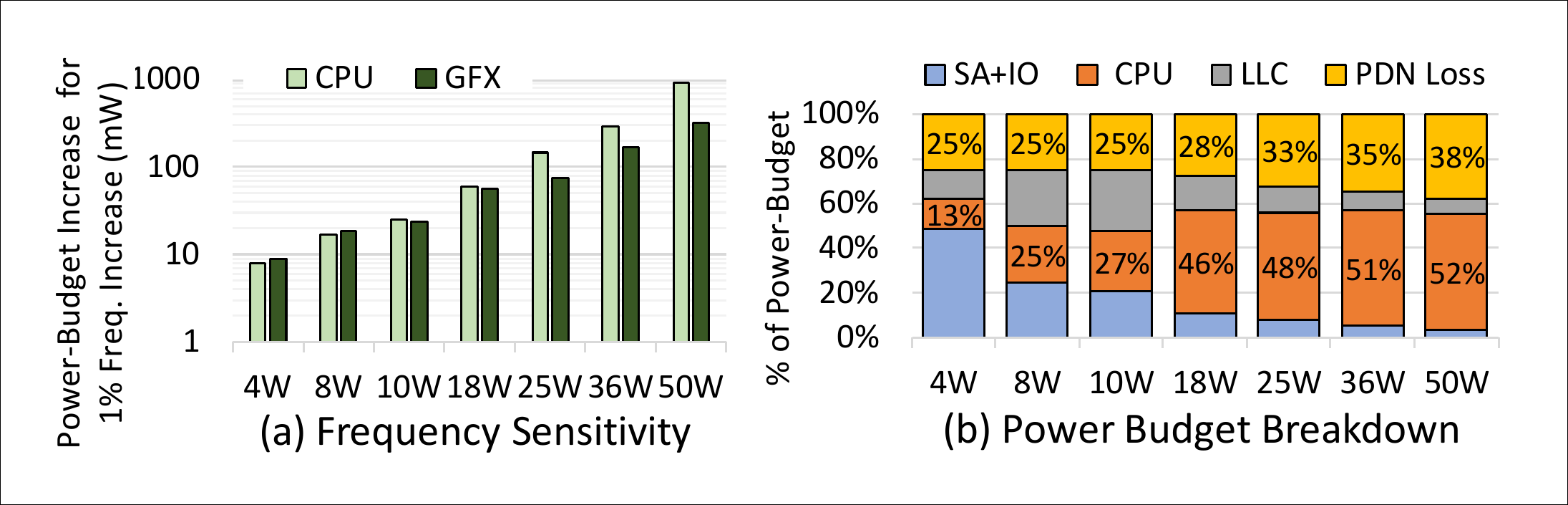}\\
   \caption{Using our performance model, we show (a) \ja{the additional power-budget required (y-axis) to increase the clock frequency of a \ja{CPU}/graphics by $1\%$ when running CPU-/graphics-intensive workloads, relative to the baseline frequency of each TDP (x-axis),} and (b) \ja{percentage (y-axis) of the total TDP power-budget (x-axis) that is allocated to CPU-cores, LLC, IO and SA, and PDN power \jkz{loss} for a CPU-intensive workload.}
   }\label{fig:perf_sens}
   \end{center}
\vspace*{-3mm}
 \end{figure}

Using our performance model, we plot in Fig.~\ref{fig:perf_sens}(a) the additional power-budget required (y-axis) to increase the clock frequency of a \ja{CPU}/graphics \jk{domain} by $1\%$ when running CPU-/graphics-intensive workloads, relative to the baseline frequency of each TDP (x-axis). \ja{We observe that, compared to a high-TDP (e.g., $50W$) processor, a low-TDP (e.g., $4W$) processor \jk{requires only a} small amount of power (e.g., ${\sim}9mW$) to increase the clock frequency of a \ja{CPU}/graphics \jk{domain} by $1\%$.}
\ja{Fig. \ref{fig:perf_sens}(b) shows the percentage (y-axis) of the total TDP power-budget (x-axis) that is allocated to the CPU-cores, LLC, IO and SA, and PDN power losses for a CPU-intensive workload (no budget is allocated to graphics in this workload). In each TDP, we use the PDN among three commonly-used PDNs (i.e., MBVR, IVR, LDO) \jk{that} \jc{\emph{maximizes}}  \jc{PDN} power loss (e.g., IVR for $4W$ and MBVR for \jkv{$50W$), to} show the effect of using an unoptimized PDN on different processor domains' power budgets.
We find that in a low-TDP processor, a relatively small fraction (e.g., only $13\%$ of a $4W$ TDP) is allocated to CPU-cores compared to a higher-TDP processor (e.g., about $52\%$ of a $50W$ TDP), while PDN power \jkz{loss is} $25\%$ or more (i.e., ETEE of $75\%$ or less).
\jk{If we use a PDN with a higher ETEE for each TDP (e.g., $5\%$ higher ETEE\jb{, which translates to $5\%$ lower PDN \jkz{power loss}}), we can increase the CPU-cores' power-budget by the spared power on PDN \jkz{loss} (e.g., $5\%$), thereby increasing the workload's performance. We illustrate the impact of a PDN's ETEE with} the following example.
}


\noindent \textbf{Impact of PDN ETEE on System Performance.} 
For a $4W$ TDP processor, the \je{domains' nominal} power \jkx{consumption (i.e., the sum of each domain's \je{nominal} power consumption}) is approximately 3W. \jkx{To find the total processor power consumption, we must account for the PDN power conversion loss by dividing the  \je{domains' nominal} power consumption by the \jkc{PDN's} ETEE. Therefore, the PDN's ETEE can dictate the amount of remaining power budget for reallocation across the domains to improve system performance. For example, we can increase the CPU-cores' clock frequency by 1\% for each \jkx{9mW increase in the CPU-cores' power budget at} a $4W$ TDP (shown in Fig.~\ref{fig:perf_sens}(a)).}

\jkx{To show how even a small difference in ETEE can have a significant impact on system performance, assume we have two PDNs: 1) $PDN_1$ with $ETEE_1$=$75\%$, and 2) \jkx{$PDN_2$} with $ETEE_2$=$80\%$. The total processor power consumption of $PDN_1$ and $PDN_2$ are $4W$ ($3W/0.75$) and $3.75W$ ($3W/0.8$), respectively. According to our model (shown in Fig.~\ref{fig:perf_sens}(a)), the additional $250mW$ ($4W-3.75W$) saved by using $PDN_2$ (instead of $PDN_1$) could be allocated to increasing the \jkc{CPU cores'} clock frequency \jd{by $28\%$}. This would increase} the performance of a \ja{highly-scalable} workload \jd{by $28\%$}.

\subsection{{\modl} Assumptions and Limitations}
\hjorg{\noindent  \textbf{Assumptions.} Our {\modl} model makes three main assumptions. 
First,  {\modl} assumes that the system operates within a thermal design power (TDP) limit. The power management unit allocates 1) a power-budget to the SA and IO domains, which have  nearly constant power consumption across different TDPs, and 2) the remaining power-budget to the compute domain (cores and graphics). The compute \jkz{domain} power-budget is divided between the cores and the graphics \jkz{engines} based on the \jkz{running workload} (e.g., CPU- versus graphics-\jkx{intensive} workload).
Second, {\modl} assumes the same routing resources for all PDNs. 
Therefore, for PDNs in which multiple domains share a single VR (e.g., IVR, LDO), the routing resources of these domains are \ja{combined}.}
Third, {\modl} assumes that all voltage emergencies are handled by both 1) existing decoupling capacitors and 2) existing architectural techniques. \ja{This is a reasonable assumption for modern client processors \cite{rotem2016system,ananthakrishnan2019controlling,singh2018zen}}.

\hjorg{\noindent  \textbf{Limitations.} Our {\modl} model has two main limitations. First, the model predicts the ETEE based on average values of inputs over a time interval (e.g., during residency in a power state). To provide the dynamic ETEE of a workload (e.g., during multiple system power states within a workload), {\modl} should be run for each time interval separately with the appropriate input for the examined time interval. \ja{However, this is not a big limitation since \jkz{doing so} can be automated (e.g., using a script) once data for multiple intervals is collected.} 
Second, the model considers the processor and the off-chip VRs as a single thermal domain (i.e., as sharing the same TDP), which is true for many systems \cite{paterna2015modeling}. However, the {\modl} model does not provide the effect of thermals on power and performance for a system in which the processor and off-chip VRs are in two different thermal domains.}

\section{{\modl} Validation}\label{sec:method}

PDNs in modern  client processors have complex designs, and they involve several components integrated on die, package, and board.
For example, the IVR design includes multiple components such as 1) buck regulator \jkz{bridges \cite{2_burton2014fivr}}, 2) control modules that generate the pulse width modulation (PWM) \jkz{signals \cite{22_hazucha2005233,perreault2009opportunities,kim2008system}} and activate IVR phases, 3)  air core inductors \jkz{(ACI) \cite{2_burton2014fivr,22_hazucha2005233}}, and 4) Metal Insulator Metal (MIM) capacitors \cite{2_burton2014fivr}. 
In addition, several IVR parameters (e.g., thresholds for voltage-regulator power-states) and algorithms (e.g., phase-shedding management) are typically configured and tuned post-silicon. Therefore, modeling these designs with, for example, \hlb{\emph{SPICE}\lois{~\cite{Nagel1973}} is inaccurate and unsuitable for validating our power models.}
Instead, we obtain the input parameters (shown in Table \ref{tab:params})  to {\modl} and validate the three power models of {\modl} \jkz{with} \emph{real} experimental data from our lab that we \lois{collect} using two different sets of \lois{benchmark} traces that are typically used to evaluate  client processors. 

In this section, we present the 1) experimental setup used to obtain {\modl} \jkz{model} parameters, 2) \lois{methodology for obtaining {\modl} model parameters}, and 3)  \lois{{\modl} validation process}.

\subsection{Experimental Setup}\label{exp_setup}

\noindent \textbf{System Setup.} To measure power and validate our power models, we use two systems with the configurations shown in Table \ref{tbl:sys_setup}. \jkz{Intel} Broadwell and Skylake architectures use IVR \cite{5_nalamalpu2015broadwell} and MBVR \cite{21_doweck2017inside} PDNs, respectively.

\begin{table}[!h]
\centering
\caption{Processor configurations and PDNs}
\label{tbl:sys_setup}
\resizebox{0.90\linewidth}{!}{

\begin{tabular}{cl}
\hline
\multirow{3}{*}{Processors}                 & \multicolumn{1}{c}{1) i7-5600U\cite{intel5600U} Broadwell architecture}                                                                                             \\ 
                                            & ~~~~PDN topology: IVR\cite{5_nalamalpu2015broadwell}                                                                                                  \\ 
                                            & 2) i7-6600U\cite{intel6600U} Skylake architecture                                                                                                \\ 
                                            & ~~~~PDN topology: MBVR \cite{21_doweck2017inside}                                                                                                                   \\
                                            & \begin{tabular}[c]{@{}l@{}}      
                                    L3 (LLC) cache: 4~MB.\\  Process technology \jkz{node}: 14~nm\end{tabular} \\ \hline
\multicolumn{1}{l}{\multirow{2}{*}{Memory}} & DDR3L-$1600~MHz$\lois{~\cite{jedec-lpddr3}}, non-ECC,                                                                                                                          \\
\multicolumn{1}{l}{}                        & dual-channel, 8~GB capacity                                                                                                                                  \\ \hline
\end{tabular}
}
\end{table}

\noindent \textbf{Benchmark Traces.}
To obtain the input parameters (shown in Table \ref{tab:params}) for our models and validate the models, we use \jkx{approximately} 5000 traces from a wide variety of benchmarks, typically used in evaluating  client processors. We use ${\sim}3000$ single threaded traces, ${{\sim}1200}$ multi-programmed traces, and ${\sim}750$ graphics traces comprising of 1) representative \ja{CPU- and graphics-intensive} workloads including SPEC CPU2006 \cite{18_SPEC}, Sunspider \cite{bench_sunspider}, PhotoShop \cite{bench_photoshop}, Illustrator\cite{bench_illustrator} SYSmark \cite{sysmark}, HandBrake \cite{bench_handbrake}, 3DMark06\cite{17_3DMARK}, Crysis \cite{bench_crysis}, 2) representative battery life workloads such as office productivity workloads (e.g., MobileMark\cite{mobilemark}), video conferencing and streaming workloads, and web-browsing workloads \cite{19_MSFT}, and 3) synthetic traces of power-virus \cite{21_doweck2017inside} for each domain\lois{,} which can be generated using tools such as McPAT\cite{li2009mcpat}, SYMPO \cite{ganesan2010system} or Intel's Blizzard~\cite{anshumali2010circuit}.

\noindent \textbf{Power Measurements.}
For the platform \emph{power measurements}, we use a Keysight N6705B DC power analyzer \cite{keysight_N6705B} equipped with an N6781A source measurement unit (SMU) \cite{keysight_acc}.
The N6705B (equipped with N6781A) accuracy is around $99.975\%$~\cite{keysight_acc}. 
The power analyzer measures and logs the instantaneous power consumption of different device components. Keysight's control and analysis software \cite{keysight_N6705B} is used for data visualization and measurement management. 
\jd{For more detail, we refer the reader to the Keysight manual \cite{keysight_N6705B} and to our prior work \cite{haj_connected_standby}}.


\subsection{Obtaining {\modl} \lois{Model}  Parameters} 
We describe the process we use to  obtain each of the input parameters to {\modl} models. A summary of the main parameters is shown in Table \ref{tab:params}.

\noindent \textbf{VR Efficiency Curves -- Input Parameters.}
We measure two sets of parameters \lois{for 1)} \emph{on-chip VR efficiency} (i.e., $\eta_{IVR}$ and $\eta_{LDO}$) and \lois{2)} \emph{off-chip VR efficiency} (i.e., $\eta_{VIN}$, $\eta_{GFX}$, $\eta_{SA}$, and $\eta_{IO}$). We perform the measurements on our systems across multiple values in the operational range of the 1) VR input voltage (e.g., 7.2V, 9V, 12V for off-chip VR; 1.6V and 1.8V for IVR), 2) VR output voltages (e.g., 0.5V, 0.6V, 0.7V, 1V, 1.8V), and 3) load current. 

\begin{figure}[hbt!]
   \begin{center}
   \includegraphics[trim=0.6cm 1.cm 0.6cm 1.cm, clip=true,width=0.95\linewidth]{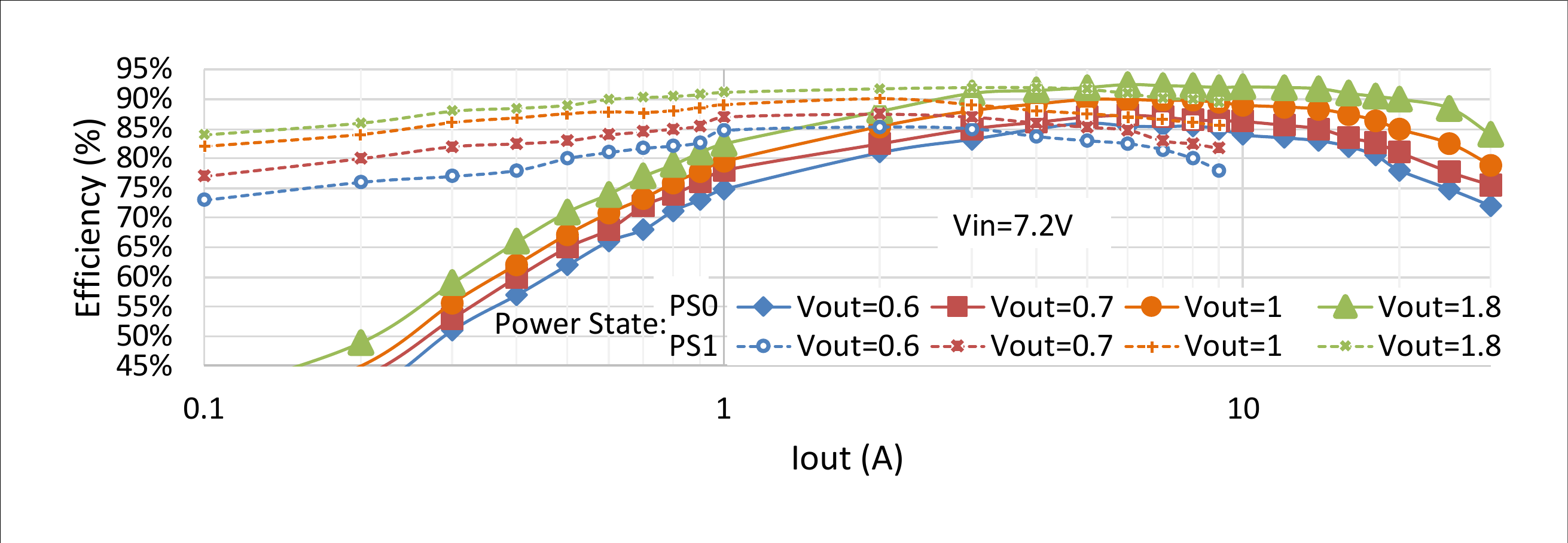}\\
   \caption{Off-chip VR efficiency curves as a function \ja{of: 1) output current (Iout, x-axis), 2) output voltage (Vout), 3) VR power-states (only PS0 and PS1 shown), and 4) input voltage (Vin, only 7.2V is shown).}}\label{fig:vr_efficiency}
   \end{center}
 \end{figure}

We measure the \emph{off-chip VR efficiency} ($\eta_{VIN}$, $\eta_{GFX}$, $\eta_{SA}$, and $\eta_{IO}$) by 1) connecting the VR input (output) to channel A (B) of the DC power analyzer, which we configure as the power supply (\jd{DC electronic load})  \cite{Keysight_N6782A}. This setup enables us to 1) measure the input and output power, and 2) sweep over the ranges of the load current, output voltage and input voltage \lois{values}, and log the data into the host PC that runs the control and analysis software. We also measure the efficiency for each VR power-state for VRs that support multiple power-states (e.g., $V_{IN}$ supports PS0, PS1, PS3 and PS4). Fig.~\ref{fig:vr_efficiency} shows the efficiency curves for the off-chip VRs (i.e.,  $V\_Core$, $V\_GFX$, $V\_SA$, $V\_IO$ and $V\_IN$) \lois{as a function of} multiple output voltages, one input voltage (7.2V) and two VR power-states (PS0 and PS1).

\begin{figure*}[b]
  \begin{center}
  \includegraphics[trim=.95cm 0.7cm 0.95cm .8cm, clip=true,width=0.95\linewidth,keepaspectratio]{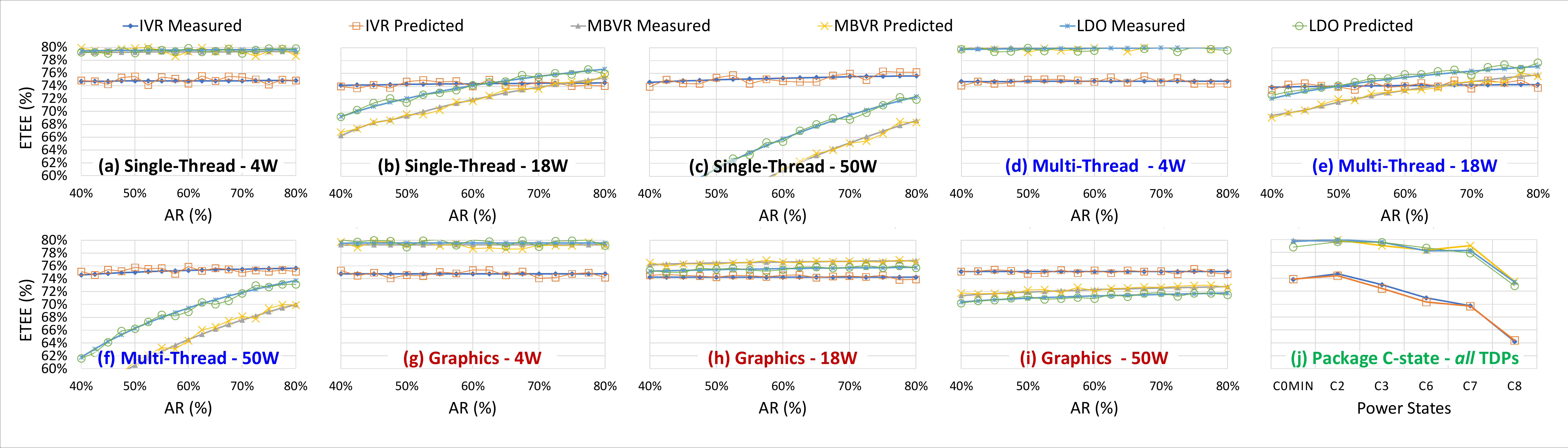}\\
  \caption{\jk{PDNspot} validation results. (a)--(i) End-to-End power-conversion  \jd{efficiency} (ETEE)  for single-threaded, multi-threaded and graphics traces at $4W$, $18W$ and $50W$ TDP with varying application ratios (AR). (j) shows the  results for battery life related power-state: C0 with minimum frequency (C0MIN) and package C-states (C2/3/6/7/8) \cite{haj2018power,haj2018energy,gough2015cpu}.}
  \label{fig:validation}    
  \end{center} 
 \vspace*{-5mm}
\end{figure*}

We measure \emph{IVR efficiency} ($\eta_{IVR}$) using the Broadwell processor. Since the IVR is integrated into the processor, it is impossible to disconnect the native load (e.g., cores, graphics engines) and connect a high current load directly to the output of \jkz{an} IVR. Therefore, to measure the IVR efficiency, we operate the processor in a special Design For Test (DFT) mode \cite{2_burton2014fivr}. We also operate the processor clock tree at varying frequencies to enable a large effective adjustable load current. We measure the current and voltage at the output and input (i.e., output of the $V_{IN}$ in Fig. \ref{fig3_three_schems}) of the IVR \cite{2_burton2014fivr}. Next, we calculate the input and output power and plot the efficiency curves as a function of load current and output voltage. Table \ref{tab:params} \jc{(On-chip VR Efficiency)} shows the range of the measured IVR efficiency \jkx{($81\%$--$88\%$)}. The actual curves in {\modl} plot the efficiency as a function of input voltage, output voltage and output current.


We measure the \emph{LDO \jkz{VR} efficiency} ($\eta_{LDO}$) in two steps. First, since the LDO \jkz{VR} is not implemented in our experimental systems, we emulate the LDO \jkz{VR} static behavior using the power-gates that exist in the MBVR PDN of the Skylake processor, a technique\footnote{By controlling the number of the conducting \jkx{power-gate transistors and their gate voltages}, the power-gate behaves like an LDO \jkz{VR}. The actual LDO \jkz{VR} implementation has additional circuitry (e.g., to handle load transient response, digital control of the LDO \jkz{VR} output).} which is used by Intel \cite{luria2016dual} to implement an LDO \jkz{VR}.
Second, we measure the input and output power of the LDO \jkz{VR} under varying load current, input and output voltages and plot the efficiency curves. The LDO \jkz{VR} efficiency is the ratio between the output and the input voltage times the ratio between input and output current (also known as current efficiency), i.e., $\eta_{LDO} = (V_{OUT}/V_{IN}) \cdot (I_{OUT}/I_{IN})$. Our measurements show that the current efficiency, i.e., ${I_{OUT} }/{I_{IN}}$, is more than $99\%$ as tabulated in Table \ref{tab:params}.

\begin{sloppypar}
\noindent \textbf{Nominal Power of Domains -- Input Parameter.}
We measure the \emph{nominal power} ($P_{NOM}$) input parameter of each domain (i.e., cores, LLC, graphics, SA, and IO) directly on the Skylake system when \jkz{running traces} of single threaded, multi-threaded and graphics workloads. We \jkz{log} the measured power of each trace and its application ratio (i.e., $AR$, \ja{discussed in Sec. \mbox{\ref{sec:vr_params}}}).
\end{sloppypar}

\noindent  \textbf{Other Input Parameters}.  We measure the \emph{Load-line impedance ($R_{LL}$)} from a domain's input to the output of the off-chip VRs for each domain directly on Skylake and Broadwell Systems. \lois{We measure \emph{peak-power}} (i.e., $P_{peak}$) when running power-virus traces. We estimate \emph{leakage-power fraction} ($F_{L}$) using a post-silicon technique, \hlb{thermal \jkz{conditioning \cite{dev2013power,hamann2006hotspot,cochran2010post}}, by 1) increasing the processor temperature while running a load with constant voltage and frequency (i.e., constant dynamic power), 2) measuring the associated changes in power consumption, and 3) extrapolating the domain's power fraction which is affected by temperature, as the leakage power depends exponentially on temperature whereas the dynamic power is not affected by temperature\ja{~\cite{haj2018power,haj2018energy,15_jakushokas2010power,gough2015cpu}.}}



\subsection{PDNspot Validation}
We validate {\modl} by comparing the predicted ETEE obtained from each {\modl} model (i.e., IVR, MBVR, and LDO) with the ETEE measurements on \emph{real} systems. 

To \lois{validate {\modl}, we use \jkz{as reference} the total power \ja{consumption} of \jd{real \jkc{Intel} processors (Broadwell, Skylake, and  Skylake with emulated LDO PDN)} \emph{measured} from the main power supply (battery/PSU) \ja{for} each of the PDNs ($P_{IVR}$, $P_{MBVR}$, and $P_{LDO}$). \ja{We use {\modl} to obtain the \emph{predicted} power consumption of each PDN.}
We use a subset \ja{($200$)} of the benchmark traces (single-thread, multi-programmed, and graphics described in \ja{Sec.} \ref{exp_setup}) that have various application ratios (AR). \ja{We calculate the measured (predicted) \jkz{ETEE}
of each PDN}  by dividing the total nominal power \ja{consumption} (i.e., PDN output power) by the  \ja{measured (predicted)} total power consumption (i.e., PDN input power). \ja{Finally, we calculate the accuracy of {\modl} by comparing the measured \jk{ETEE to} the predicted ETEE of each PDN.}} 


\begin{sloppypar}
We find that our three IVR, MBVR and LDO \jkz{PDN} models \jkz{in} {\modl} have an average (min/max) \emph{accuracy} of $99.1\%$ ($98.7\%$/$99.3\%$), $99.4\%$ ($98.9\%$/$99.7\%$), and $99.2\%$ ($98.6\%$/$99.6\%$), respectively, \lois{across all our \ja{$200$} workloads}.
Fig. \ref{fig:validation}(a--i) shows the validation results (measured vs. predicted \jkz{ETEEs})  for $4W$, $18W$, $50W$ TDPs when running single-threaded, multi-programmed, and graphics traces with an AR between $40\%$ to $80\%$. Fig. \ref{fig:validation}(j) shows the results for the battery life related power-state\ja{s}: C0 with minimum frequency ($C0_{MIN}$) and package C-states (C2/3/6/7/8) \cite{haj2018power,haj2018energy,gough2015cpu}. 
\end{sloppypar}





\section{Motivation: \jkz{PDN Inefficiencies in Client Processors}} \label{sec:ourmotivation}
This section makes three key \jk{empirical} observations about the three most commonly-used PDN architectures (i.e., IVR~\cite{2_burton2014fivr,5_nalamalpu2015broadwell,icelake2020}, MBVR~\cite{9_rotem2011power,jahagirdar2012power,11_fayneh20164}, LDO~\cite{singh20173,singh2018zen,burd2019zeppelin,beck2018zeppelin,toprak20145,sinkar2013low}) in modern \jk{high-end} client processors to motivate the need for a hybrid and adaptive PDN that \jkz{leverages} the advantages of each one of the three PDN architectures.

We use our validated model, {\modl}, to \jk{evaluate the efficiency of the three PDNs}. We estimate 
the off-chip current consumption, ETEE with breakdown into multiple sources of power-conversion losses, and average power consumption of a processor using each of the three PDNs. 
We use \jk{a total of} \ja{300} \ja{CPU}-intensive, graphics-intensive, and video playback \jk{workload} traces to evaluate each PDN. 

\begin{figure}[h!]
\begin{center}
  \includegraphics[trim=.9cm .7cm .9cm .7cm,clip=true,width=0.9\linewidth]{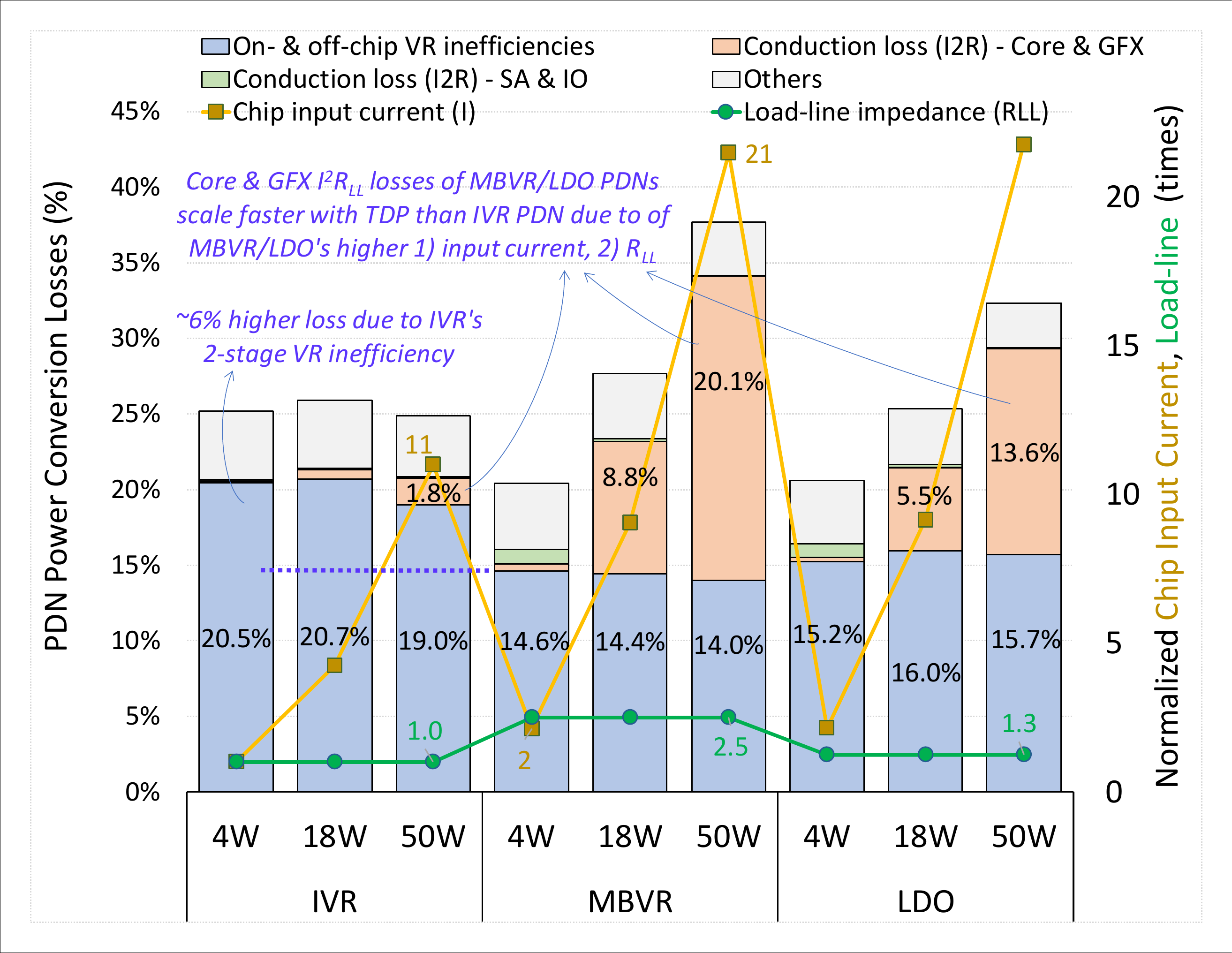}\\
  \caption{Breakdown of the power \ja{conversion} loss of the three PDNs when running a CPU-intensive workload ($AR$=$56\%$) at $4W$, $18W$, and $50W$ TDPs. \ja{Conduction loss ($I^2R$) and \jd{on-chip} \& off-chip VR infficiencies are the most prominent losses}. \jk{Normalized (to IVR PDN) \jkz{chip} input} current \ja{(\je{$I$, i.e., }from off-chip VRs)} and load-line \ja{impedance} ($R_{LL}$) are shown as line plots.}
  \label{fig:pdns_compare1}
  \end{center}
\vspace*{-3mm}
\end{figure}

Based on our evaluation results shown in Figures \ref{fig:validation} and \ref{fig:pdns_compare1}, we make three key observations.

\noindent \textbf{Observation 1.} We observe that when \jkz{executing} CPU- and graphics-intensive workloads, the \text{IVR} PDN has a \emph{lower} ETEE \jkz{at the $4W$ TDP (Figures \ref{fig:validation}.a,d,g) and a \emph{higher} ETEE at the $50W$ TDP (Figures \ref{fig:validation}.c,f,i) compared to \text{MBVR} and \text{LDO} \jkz{PDNs} across the entire range of \jc{tested} ARs}. The ETEE \emph{\jk{crossover} point}\ja{, \jkz{at} which the IVR ETEE \jkz{becomes} higher than the MBVR/LDO ETEE,} \jk{exists} at some TDP  between $4W$ and $50W$.

Fig. \ref{fig:pdns_compare1} provides more insight into this observation \jkz{with breakdowns of PDN power conversion loss. We find that at $4W$ TDP, the dominating contributor to the PDN power conversion loss are the \jd{on-chip} and off-chip VR inefficiencies. At $4W$ TDP, the IVR PDN has a lower ETEE than the MBVR and LDO PDNs due to the higher power conversion inefficiencies of \jkc{the IVR PDN's on-chip and off-chip VRs}. At a $50W$ TDP, we find that MBVR and LDO PDNs have lower ETEEs due to their high $I^2R$ loss in \jk{core} and graphics domains. The high $I^2R$ loss is due to:} 1) a ${\sim}2\times$ higher \jc{chip input} current in the MBVR and LDO PDNs compared to \jkz{the IVR PDN}\footnote{The IVR PDN reduces the chip \jc{input} current because it uses high input voltage from \jkz{the} first-stage VR into the chip (Sec. \ref{sec:background}).}, 
and 2) a  $2.5{\times}/1.3\times$ higher load-line \jc{impedance} ($R_{LL}$) in the MBVR/LDO PDNs compared to the IVR PDN\footnote{\hlo{The IVR and LDO PDNs have lower $R_{LL}$ compared to MBVR because both IVR and LDO PDNs share routing resources from external VRs into the chip's package and die.}}.
We conclude that the MBVR and LDO PDNs are more \jkz{efficient at} a low TDP (e.g., $4W$) compared to the IVR PDN, while the IVR PDN is more efficient \jk{at a} high TDP (e.g., $50W$). 





\noindent \textbf{Observation 2.}
We observe that the PDN ETEE is affected not only by the TDP (as discussed in Observation 1) but also by the workload's \jk{Application Ratio (AR)} and the workload type, i.e., single-threaded, multi-threaded, and graphics. 

Fig. \ref{fig:validation}(a--i) shows that the MBVR and LDO PDN ETEEs \ja{increases} with AR, which is most \jkz{pronounced at} 18W and $50W$ TDPs. This phenomenon is due to the load-line (described on Sec.~\ref{sec:background})\jk{, which results in a lower voltage-guardband when running workloads with higher ARs.}

Fig. \ref{fig:validation}(b,e,h) show that the single-thread, multi-thread, and graphics \jk{workloads} (all at \jk{the} same TDP of 18W) have different ETEE curves. 
For example, for the graphics workload in Fig. \ref{fig:validation}(h), the IVR PDN is less efficient than the other two PDNs for the  entire AR range (with a \jk{crossover} point around 21W TDP, not shown in Fig. \ref{fig:validation}\ja{, \jkz{at} which the IVR ETEE \jkz{becomes} higher than the MBVR/LDO ETEE}), while the other two \ja{workloads} have \jk{crossover} points at different ARs within \jk{the} 18W TDP. 

Fig. \ref{fig:validation}\ja{(a--f)} shows that \jk{the LDO ETEE is higher than the MBVR ETEE for CPU-intensive} (single- and multi-threaded) workloads, but is lower than the MBVR ETEE for graphics-intensive \jk{workloads}. Note that the LDO inefficiency is more dominant in graphics workloads, due to the high voltage difference between the \jk{core} and graphics domains because in graphics-intensive workloads, the graphics-engine runs at relatively high frequencies (and voltages) while cores are kept at low frequencies (and voltages). 
\ja{Therefore, the LDO PDN 1) sets the off-chip (i.e., first stage) VR voltage to the high voltage level required by the graphics-engines (e.g., $0.9V$) while activating the graphics-engines' on-chip LDO (i.e., \jkz{second-stage}) VR in bypass-mode, and 2) uses the core's on-chip LDO (i.e., \jkz{second-stage}) VR to regulate the voltage down to the low voltage level required by the core (e.g., $0.5V$). Doing so, results in very low power conversion efficiency of the core's \jkz{LDO VR} (e.g., ${\sim}0.5/0.9=55\%$, as discussed in Sec. \ref{Buck_Converters}), thereby reducing the ETEE of the LDO PDN. }

We conclude that, in addition to the TDP, the AR and workload type have significant effects on \jk{each} PDN's ETEE. \ja{Particularly, lowering the workload's AR degrades the ETEE of MBVR and LDO PDNs due to load-line effect, while using graphics\ja{-intensive} \jkz{workloads reduce} the LDO ETEE compared to \ja{CPU-intensive} workloads due to the high voltage requirement difference between \jkz{the} core and graphics domains.} 

\noindent \textbf{Observation 3.}
We observe that the ETEE of \jk{the} IVR PDN is significantly lower than that of MBVR and LDO \jkz{PDNs} for \ja{computationally} light workloads \ja{(e.g., video playback, web browsing, office productivity applications \cite{mobilemark,19_MSFT,sysmark})} and low-power states across \emph{all} TDPs.
Fig. \ref{fig:validation}(j) shows the ETEE of the three PDNs in 1) $C0_{MIN}$, an active power-state in which the \jk{core and graphics domains operate at} their \emph{lowest} frequencies, and  2) package C-states (C2, C3, C6, C7, and C8 \cite{haj2018power,haj2018energy,gough2015cpu}), low power-states of the processor. The processor uses these power-states, for \emph{all} TDPs, to reduce energy consumption (thereby increasing battery life of battery-powered devices) when the processor runs a light (i.e., low computational intensity) workload or once the processor is partially/fully idle. We explain the effects of ETEE in these power-states on battery life using a video playback workload example.


The video playback \cite{19_MSFT} workload is a \ja{computationally} \emph{light} workload \jk{that} operates in three main power-states during each video-frame. 
First, a $C0_{MIN}$ power-state\jk{,} which consumes $P_{C0_{MIN}}$=$2.5W$ nominal power for $R_{C0_{MIN}}$=$10\%$ ($R_{C0_{MIN}}$ is the residency of power state ${C0_{MIN}}$ \jk{in terms of the fraction of execution time}) of the frame's time. In this state, the cores and graphics engines prepare a video-frame and store it in main memory.
Second, a $C2$ power-state, which consumes $P_{C2}$=$1.2W$ nominal power for $R_{C2}$=$5\%$ of the  frame's time. The cores and graphics engines are idle (power-gated) in this state. In $C2$, the display-controller fetches part of the frame from main memory into a local buffer inside the display controller.    
Third, a $C8$ power-state, which consumes $P_{C8}$=$0.13W$ nominal power for $R_{C8}$=$85\%$ of the frame's time. In $C8$, the display controller reads frame data from its local buffer and displays it \jk{on} the display panel, while the rest of the processor is idle (e.g., main memory \jkz{is} in self-refresh). We calculate the average power of the video playback workload by summing the fractional power of each power-state taking into account the ETEE in each state (denoted by $\eta_{C0_{MIN},2,8}$). Hence, the average power is given by: $P_{C0_{MIN}}\cdot R_{C0_{MIN}}/\eta_{C0_{MIN}} + P_{C2}\cdot R_{C2}/\eta_{C2} +  P_{C8}\cdot R_{C8}/\eta_{C8}$. The video playback average-power results \ja{(shown in Fig. \ref{fig:all_flexwatts_res}(c))} \jkz{show} that MBVR and LDO PDNs have \ja{${12\%}$} and ${11\%}$ \emph{lower} average power\ja{, respectively,} than the IVR PDN.
We conclude that the IVR PDN is energy-inefficient \jd{for} \jk{computationally-}light \jkz{workloads,} which \jk{negatively impacts both} energy consumption and battery life.

\noindent \textbf{Summary.} We conclude that there is no single PDN for modern client processors that maintains a high ETEE across all TDPs, workload types and application ratios (ARs).
These observations motivate us to build a \emph{hybrid} and \emph{adaptive} PDN that utilizes the advantages of each one of the three PDN architectures, as we describe in Sec. \ref{sec:new_pdn}. 

\vspace*{3mm}
\section{{\tech} }\label{sec:new_pdn}

\begin{figure}[h]
  \begin{center}
  \includegraphics[trim=.6cm .6cm .6cm .60cm, clip=true,width=0.95\linewidth,keepaspectratio]{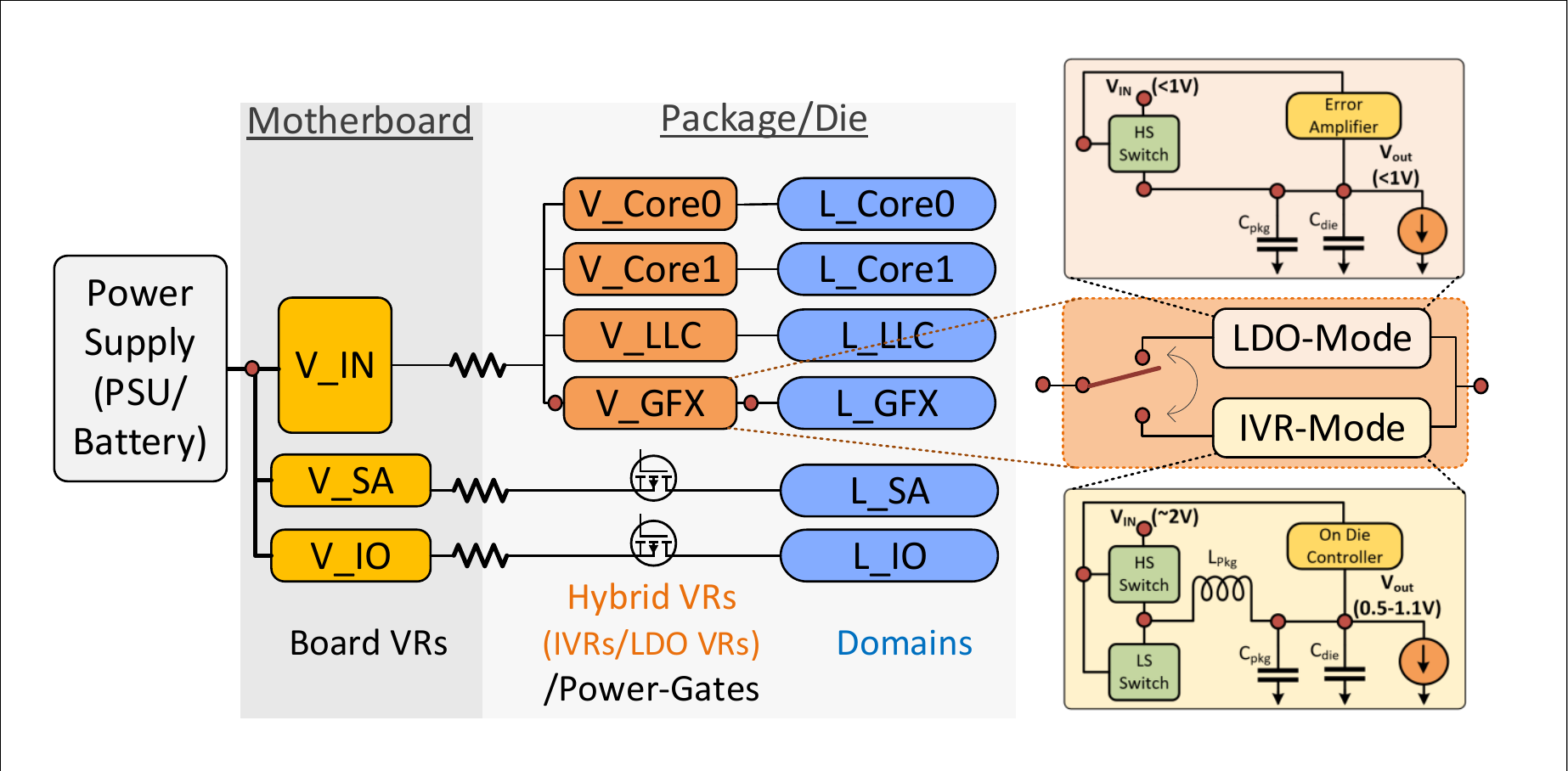}\\
  \caption{Our hybrid adaptive PDN ({\tech}). {\tech} uses \ja{an off-chip VR to each system domain with a \jb{low and} narrow power consumption range (i.e., SA and IO domains)}. 
  \ja{For system domains with a wide power consumption range (e.g., CPU cores and graphics engines), {\tech} allocates a hybrid PDN. This hybrid PDN \jk{can dynamically switch} between two modes, \texttt{IVR-Mode} and \texttt{LDO-Mode}, based on the \jkz{expected ETEE benefits} of each mode for the current workload and power consumption.} 
 \ja{The hybrid PDN} shares \jkz{between IVR and LDO modes} 1) on-chip resources such as the high-side (HS) NMOS power switch in the IVR PDN as illustrated on the right side, and 2) off-chip VRs ($V\_IN$).}
 \label{fig:pdn_mil}  
  \end{center}
\vspace*{-3mm}
\end{figure}

\sr{We} present \emph{\tech}, a hybrid adaptive PDN for modern processors that maintains a high ETEE for the wide power \sr{consumption} range and \sr{workload \jkz{diversity}} of client processors. {\tech} is based on \emph{three} key ideas.
First, \sr{it} combines IVRs and LDOs in a novel way \jk{to share} multiple \jkc{on-chip and off-chip} resources and \jd{thus} reduce \jd{BOM, as well as board and die area overheads}, as illustrated in Fig. \ref{fig:pdn_mil}. 
This hybrid PDN \jk{is allocated for processor domains with a wide power consumption range (e.g., CPU cores and graphics engines) and \jkz{it} dynamically switches between two modes,} \texttt{IVR-Mode} and \texttt{LDO-Mode}\sr{, based on the efficiency of each mode,} using a special power-management flow.
Second, {\tech} statically allocates \sr{an} off-chip VR to \sr{each} \ja{system} domain with a \jb{low and} narrow power \sr{consumption} range (i.e., SA and IO \sr{domains}). \sr{This is because} unlike in \sr{compute} domains, the power consumption of the system-agent (SA) and IO domains does \emph{not} significantly scale with TDP (as \sr{shown} earlier in Fig. \ref{fig:perf_sens}(b)) or workload's AR. Thus, \jkz{it is} more energy-efficient to place \jkz{each of} them on a dedicated off-chip VR compared to using \jkz{an} on-chip VR\footnote{\jkc{AMD uses the same strategy for their LDO PDNs} \cite{singh2018zen} (Fig. \ref{fig3_three_schems}(c))}.
Third, {\tech} introduces a new prediction algorithm that automatically determines which PDN mode (\texttt{IVR-Mode} or \texttt{LDO-Mode}) would be \sr{the} most beneficial based on system and workload characteristics.  
For example, {\tech} can operate \sr{in} \texttt{LDO-Mode} (\texttt{IVR-Mode}) when the processor runs a light (heavy) workload such as  video playback (Turbo Boost), or when the processor operate\sr{s} \sr{at} low (high) TDP such as $4W$ ($50W$). \jkz{{\tech}} uses a runtime ETEE prediction algorithm to select the operation mode (i.e., \texttt{LDO-Mode}  or \texttt{IVR-Mode}) \sr{that} maximizes ETEE.

\noindent \textbf{\hlo{Hybrid PDN and Resource Sharing.}}
\ja{We build the {\tech} PDN by modifying a baseline IVR PDN, shown in Fig. \ref{fig3_three_schems}(a), in two ways.}
\ja{First, we replace the two on-chip IVRs of the SA and IO domains (i.e., V\_SA and V\_IO IVRs) \jk{with} two off-chip VRs and two on-chip power-gates, as illustrated in Fig. \ref{fig:pdn_mil}.}
\ja{Second, we \jc{implement \emph{hybrid VRs}\jd{,} which} extend each of the remaining IVRs (i.e., \jb{V\_Core0/1, V\_LLC and V\_GFX IVRs in Fig. \ref{fig3_three_schems}(a)}) by implementing an LDO \jc{VR} using the existing resources of the IVR, as illustrated \jk{in Fig.} \ref{fig:pdn_mil} (right side).}
\hjov{
By doing so, we \sr{enable} \jk{a} \ja{\emph{hybrid PDN} that has} two modes of operation, \texttt{IVR-Mode} and \texttt{LDO-Mode}, with low cost and \jkz{low} area overhead.
As illustrated in Fig. \mbox{\ref{fig:pdn_mil}}, \jc{each} \jkz{hybrid} \jc{VR} \jkz{shares between the two modes} \ja{1)} on chip resources such as \sr{the} high-side (HS) NMOS power switch \mbox{\cite{2_burton2014fivr}}, and decoupling capacitors (both on package and \sr{on} die)\ja{ of the baseline on-chip IVR, and 2) off-chip \jk{VRs} (i.e., $V\_IN$)}. We \sr{use} the HS power-switch to implement the LDO \jc{VR}, similar to Luria \textit{et al.}~\cite{luria2016dual}, a work carried out \sr{by} Intel \sr{that} utilizes the power-gate's power-switch to implement an LDO \jc{VR}.} 
\hjov{This architecture enables both PDN modes to share routing resources and \jk{the} power grid across board, package, and die \jkc{during operation, as illustrated in Fig. \mbox{\ref{fig:pdn_mil}}}.}

\begin{sloppypar}
\noindent \textbf{Voltage Noise-Free Mode-Switching.}
{\tech} mode-switching transitions the hybrid PDN between two mode\sr{s} (\texttt{IVR-Mode} and \texttt{LDO-Mode}). Carrying out the mode-switching while the compute domains are active may introduce \emph{voltage noise} because the two modes have very different operation principles.
In \texttt{IVR-Mode}, the off-chip VR ($V\_{IN}$) is set to a relatively high-voltage (e.g., $1.8V$) and the on-chip IVRs regula\sr{te} the voltage to the level the domain needs (e.g., $0.6V$--$1.1V$). In \texttt{LDO-Mode}, $V\_{IN}$ voltage is set to the \emph{maximum} voltage required by all domains (e.g., $0.6V$--$1.1V$) and the on-chip LDOs regulate this maximum voltage to the level the domain needs. Therefore, the mode-switching should configure the \jd{on-chip} and off-chip VRs and change their voltage level\sr{s} while \jk{transitioning} from \sr{one} mode to the other.
\end{sloppypar}

To prevent any \emph{voltage noise} during mode-switching, \jk{{\tech} performs} mode-switching while the comput\sr{e} domains are \emph{idle}. To do so, we 1) place the processor in an idle power-state for a short period, 2)  configure the hybrid PDN and update the \jd{on-chip} and off-chip VR levels, and 3) exit the idle power-state and resume the processor with the new PDN mode.
To this end, we utilize a power-management flow that places the processor into \sr{the} idle power-state, (\sr{which} exists in most modern processors \cite{21_doweck2017inside,huang2015measurement,hammarlund2014haswell,tu2015atom,gough2015cpu,haj2018power,haj2018energy,haj_connected_standby,haj2013connected}), in which the cores, LLC, and graphics \sr{units} are \jkb{turned off after their contexts are} saved into a dedicated SRAM. We leverage the C6 package C-state power management firmware flow \cite{haj_connected_standby} to implement {\tech}'s mode-switching transition flow.
{\tech} takes the following three steps to switch \jkb{between two PDN modes}.

First, the power management unit (PMU) places the \ja{system} into \jd{the} package C6 \jkz{idle power state} during which the PMU saves the context\footnote{\jk{The context} is stored into dedicated SRAMs\sr{, using} power \sr{from} an always-on VR (not shown in Fig. \ref{fig3_three_schems}) that retains \jkz{the dedicated SRAMs' contents in idle states} \cite{haj_connected_standby,haj2013connected}.} of \ja{the hybrid PDN domains (i.e.,} the \ja{CPU} cores, LLC, and \jkb{graphics)} and  \jkz{turns off} their clock and voltage.   
Second, the PMU performs the actual \jkz{mode switching} actions \ja{of the hybrid  PDN} by 
1) adjusting the $V\_{IN}$ VR voltage to a level suitable for the new mode (i.e., $1.8V$ for \texttt{IVR-Mode}, or $0.6V$--$1.1V$ for \texttt{LDO-Mode}), \jk{and} 2) configur\sr{ing} the hybrid  \jd{VRs} to operate in the new mode (as illustrate\sr{d} in Fig. \ref{fig:pdn_mil}). Third, the PMU exits the package C6 idle power-state \sr{and switches to} the active state. Doing so allow\sr{s} the processor to \jkb{resume execution} while the \ja{hybrid PDN domains}
use the new PDN mode.

\noindent \textbf{Runtime PDN Mode-Prediction Algorithm.}
\sr{So far, }we explain\sr{ed} how to switch between two PDN modes (i.e., mode-switching flow) without \sr{describing} \emph{when} to switch. {\tech} relies on our \sr{new} runtime mode-prediction algorithm whose goal is to predict which PDN mode, among the two modes, \texttt{IVR-Mode} and \texttt{LDO-Mode}, provides the best end-to-end power-conversion efficiency (ETEE).

\hjg{As shown in Fig. \ref{fig:validation}, ETEE is a function of 1) the AR and  the workload type (i.e., single-thread, multi-thread, and graphics), and 2) the TDP and the power-state of the system.
ETEE depends on the AR due to the load-line effect (discussed in Sec. \mbox{\ref{sec:vr_params}}) and shown in Equation \mbox{\ref{eqn:VDLL}}. 
The workload type affects ETEE because each of the three workload types stresses the underlying power delivery \sr{network} differently, as explain\sr{ed} in Sec. \mbox{\ref{sec:models}}.}

\sr{Algorithm \mbox{\ref{alg:pdn_switch}} depicts o}ur \jkz{mode prediction} algorithm.
\hjg{The \emph{key} idea \sr{of} our algorithm is two-fold. 
First, we store two sets of ETEE curves \jk{inside the PMU firmware, one set for \sr{the} IVR PDN and \sr{the} other set} for the LDO PDN. 
A PDN ETEE curve set is a multidimensional table\footnote{A modern PMU implements multiple curves (as tables) such as leakage power as function of temperature and voltage, voltage as function of frequency, VR power-conversion efficiency as a function of input-voltage, output-voltage and output-current\cite{haj2018power,haj2018energy,gough2015cpu,rotem2012power,rotem2015intel}.} that includes an ETEE curve corresponding to a TDP for each workload type (i.e., three curves for each TDP point). Each ETEE curve stores the ETEE values as a function of the AR (as show\sr{n} in Fig. \mbox{\ref{fig:validation}(a-i)}). We also include one ETEE curve for \jkz{power states} (as show\sr{n} in Fig. \mbox{\ref{fig:validation}(j)}).
Second, for every \emph{evaluation interval} (e.g., 10ms), we estimate each of the algorithm's input parameters (i.e., TDP, AR, workload type, and power-state). We use the estimated parameters to access the corresponding ETEE curve to obtain the ETEE values for both \mbox{\text{IVR}}-mode and \mbox{\text{LDO}}-mode. The algorithm chooses the mode that maximizes the ETEE.}
\hlp{Next, we explain how we estimate the inputs to our algorithm (i.e., TDP, AR, workload type, and power-state) at runtime.}

\begin{algorithm}[h]
\caption{{\tech} \jkz{Mode Prediction} \sr{Algorithm} }\label{alg:pdn_switch}
\footnotesize
\begin{algorithmic}[1]
\Procedure{Determine\_{\tech}\_Mode}{}
\State \textbf{Input}: TDP, AR, WL\_TYPE, PS /*power-state*/ 
\State \textbf{Output}: PDN\_Mode (\texttt{IVR-Mode} or \texttt{LDO-Mode})
\State IVR\_ETEE = \sr{estimate}\_IVR\_ETEE (TDP,AR,WL\_TYPE,PS)
\State LDO\_ETEE = \sr{estimate}\_LDO\_ETEE (TDP,AR,WL\_TYPE,PS)
\State \textbf{if} {$IVR\_ETEE \geq LDO\_ETEE$} 
\State ~~~~~~~~\Return \texttt{IVR-Mode}
\State \textbf{else} \Return \texttt{LDO-Mode}
\EndProcedure
\end{algorithmic}
\end{algorithm}

\noindent \textbf{\hlp{Runtime Estimation of the Algorithm Inputs.}}
\hjg{The PMU of a modern processor uses the TDP, AR, workload-type, and power-state in multiple power management algorithms such as 1) power-budget management (PBM) \jkz{algorithm \cite{rotem2012power,21_doweck2017inside,david2010rapl}}, 2) Turbo Boost \jkz{algorithm \cite{rotem2012power,21_doweck2017inside,rotem2015intel}}, and 3) system maximum current \jkz{protection \cite{rotem2016system,ananthakrishnan2019controlling}}.} 

\jk{The runtime-configured TDP value is available} to the PMU \cite{cTDP,cTDP2}.
\jk{To estimate} \hjg{the AR, the PMU uses activity \jkz{sensors \cite{ananthakrishnan2019controlling,vogman2018method,ardanaz2018hierarchical,shrall2017controlling,rotem2016system,linda2014dynamic,burns2016method,fetzer2015managing}} that are implemented \sr{in} multiple domains of the \sr{Intel} Skylake \jd{processor}\jkz{~\cite{shrall2017controlling,rotem2016system,burns2016method,21_doweck2017inside}}. These activity sensors estimate \sr{each} domain's activity using internal events in each domain, such as active execution ports in the core, memory stalls, type of instructions being executed (e.g., scalar, vector instructions of 128-bits/256-bits/512-bits). A dedicated weight is associated \jkz{with} each event, and the weighted sum of the events in a domain is  \jb{periodically (e.g., every millisecond)} sent to the PMU. The  weights of the activity sensors are calibrated post-silicon to \jkz{provide} a proxy \ja{of the AR}.}

\hjg{The PMU estimates the \emph{workload-type} (\mbox{\text{WL\_TYPE}}) based on the power-state (i.e., active/idle) of the cores and graphics engines. For example, if the graphics engines are active, then the workload-type is set to graphics, while if more than one core is active and the graphics \sr{engines are} 
idle\sr{,} then it is set to multi-threaded.} 

\hjg{The \emph{power-state}, i.e., package power-state, of the processor is known to PMU firmware \ja{as the PMU carries out the transitions from one package C-state to \jkz{another~\cite{haj2018power,haj2018energy,gough2015cpu}}}}.

\noindent \textbf{{\tech} Overhead.}
We estimate the latency of our {\tech} \jkz{\emph{mode switching}} flow \jk{with techniques used by} previous works that estimate the package C-state latencies \cite{schone2015wake,schone2019energy}. 
We find that 1) placing the processor into package C6 \sr{power state} takes $45{\mu}s$ (without voltage changes), 2) adjusting the \jkc{on-chip and off-chip} VR voltage levels (\jkc{assuming a \ja{latency of} ${\leq}2{\mu}s$ for on-chip VRs \cite{2_burton2014fivr,  luria2016dual}, and a slew rate of $50~mV/{\mu}s$ \cite{3_coorporation2009intel} for off-chip VRs}) takes $19{\mu}s$, and 3) exiting \sr{the} C6 power state takes about $30{\mu}s$. 
Hence, the overall flow takes near\sr{ly} $94{\mu}s$. It should be noted that the DVFS (P-state) latency \sr{on} Intel processors can take up to $500{\mu}s$ \cite{hackenberg2015energy,gough2015cpu,huang2015measurement,mazouz2014evaluation} depending on the processor's internal state, which \sr{shows that the} {\tech} flow latency \sr{is} within \sr{an} acceptable range. 


The \emph{area overhead} of {\tech} over the IVR PDN is minimal. The additional area required to implement the LDO mode using the IVR resources (i.e., the high-side NMOS power switch) is around  $0.041mm^2$ \cite{luria2016dual} at 14nm process technology \jkz{node}. \sr{This} corresponds to \sr{only} $0.04\%$ and $0.03\%$ of \sr{the} Intel dual and quad core client die \sr{sizes} \cite{Skylake_die}, respectively.     

\section{Experimental Results}
\label{sec:results}

We evaluate {\tech} with respect to  performance, battery life, board area and \jkc{bill of materials \lo{(BOM)},} compared to the three commonly-used state-of-the-art PDNs in \lo{modern} processors: IVR, MBVR, and LDO. \hjb{We also include a comparison with a \jkc{hybrid} PDN (used in \jkc{Intel} Skylake-X processors~\cite{skylakex}) that combines IVR and MBVR PDNs, which we refer to as \emph{I+MBVR}.
\jkc{Similar to the LDO PDN, I+MBVR uses off-chip VRs for the SA and IO \lon{domains}} \jc{and similar to the IVR PDN, it uses IVRs for the other domains.}} 
We evaluate the PDNs using our \lo{new} {\modl} framework described in Sec. \ref{sec:model}.



\subsection{\jc{CPU} and Graphics Performance}

We evaluate the performance of {\tech} compared to other PDN architectures (IVR, MBVR, LDO, I+MBVR), under the following scenarios: 
\begin{itemize}
\item When running SPEC \jc{CPU2006}~\cite{18_SPEC} core performance benchmarks, \jkc{on} processors with $4W$ TDP. We also show the average performance of SPEC \jc{CPU2006} as TDP varies between $4W$ and $50W$. 
\item \jk{When running 3DMark06\cite{17_3DMARK} graphics performance workloads, as TDP varies between $4W$ and $50W$.}
\end{itemize}

\jkc{We evaluate the performance of CPU- and graphics-intensive workloads \jkc{assuming a fan-less system}\footnote{\je{The junction temperature ($T_{j}$) of a fan-less small form factor device (e.g., smartphone, tablet) is typically limited by the outer surface temperature of the device~\cite{rotem2013power,xie2014therminator}}.}. \jkc{Therefore, we use a} \ja{junction temperature ($T_j$) of $80^{\circ}C$} for TDPs \lo{between} $4$--$8W$ and $100^{\circ}C$} for TDPs higher than $8W$. 

\noindent \textbf{SPEC \jc{CPU2006} \jkx{Benchmarks} at 4W TDP.}
We evaluate SPEC \jc{CPU2006}~\cite{18_SPEC} benchmarks with the maximum allowed frequency (i.e., 0.9GHz) \jkc{for a $4W$ TDP system}. For these benchmarks, \jkc{the two cores} run at the same frequency and \lo{voltage,} as in all recent client processors~\cite{2_burton2014fivr,5_nalamalpu2015broadwell,9_rotem2011power,jahagirdar2012power,11_fayneh20164,singh20173}. In addition, the voltage design point for the LLC \jkc{matches} the core voltage domain as described in Rotem \textit{et al.} \cite{rotem2009multiple}.
\jkc{Thus, the core0, core1, and LLC domains} have nearly the same voltage requirements (\jkz{\lo{except for} voltage variations \lo{due to}} manufacturing process \lo{variation}).  

\begin{figure}[!h]
  \begin{center}
  \includegraphics[trim=.95cm .7cm .8cm .8cm, clip=true,width=\linewidth,keepaspectratio]{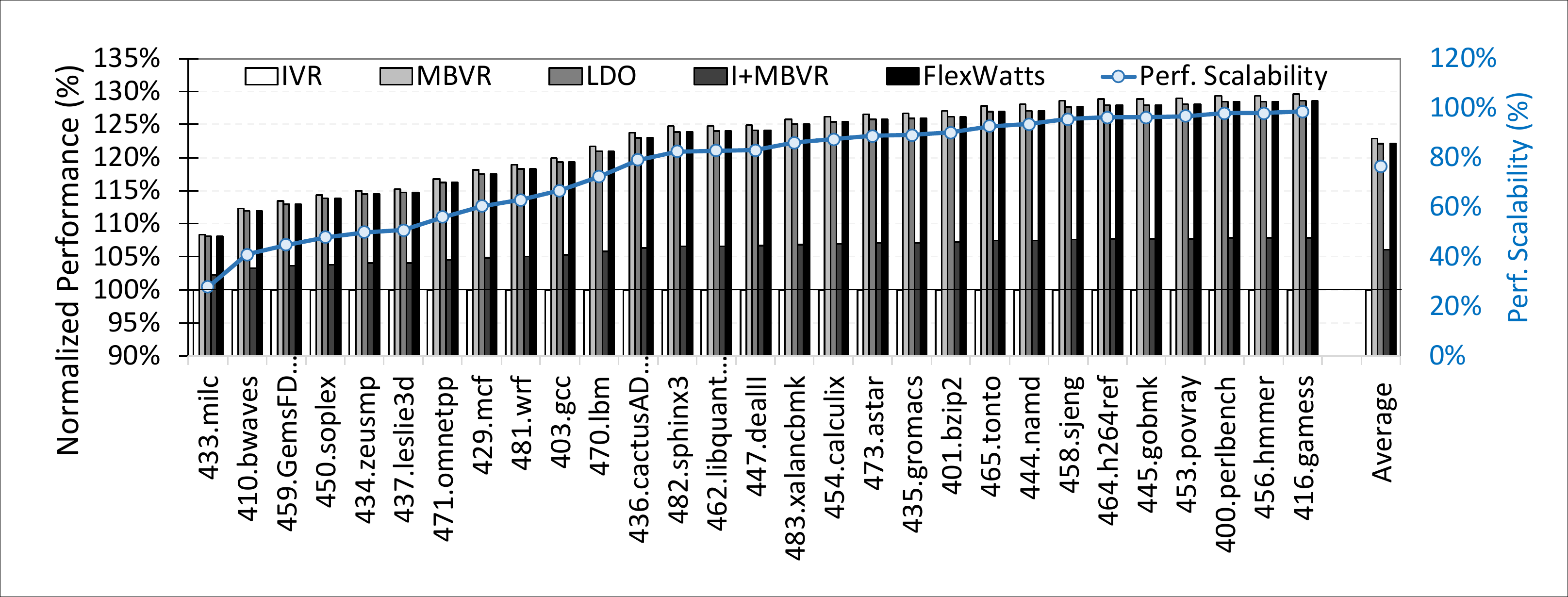}\\
  \caption{SPEC CPU2006 performance (\jkz{normalized to the IVR PDN}) \jc{with} \jkc{five} PDNs \jc{at} $4W$ TDP\jc{,} \hja{sorted (in ascending order) by the average performance-scalability of each benchmark.}  }\label{fig:perf_gain_25W_TDP}   
  \end{center} 
\vspace*{-3mm}
\end{figure}

Fig. \ref{fig:perf_gain_25W_TDP} \jkz{plots \lo{the}} performance \lo{improvement} \jc{(normalized to \lo{that of the} IVR PDN \jd{at $100\%$})} \jkz{of each SPEC CPU2006 benchmark when using each of} the \ja{five} PDNs in a \jkz{$4W$ TDP} system. {\modl} uses the \hjz{performance-scalability metric} of the SPEC CPU2006 benchmarks to estimate performance \jc{(as we discuss in Sec.~\ref{sec:perf_model})}. \hjp{Based on Fig. \mbox{\ref{fig:perf_gain_25W_TDP}}, we make \lo{four} key observations.
1) \jkz{The performance \lo{improvement} of MBVR, LDO, and {\tech}, averaged across all benchmarks, is greater than \lo{$22\%$} for \jc{the} $4W$ TDP system.} This is because MBVR, LDO, and {\tech} (which mainly operates in \texttt{LDO-Mode} at $4W$ TDP) \jkc{each} have \jkz{a higher} ETEE than IVR at low TDP. At \jkc{low TDPs}, IVR has \jkc{a larger power conversion loss} due to the two-stage (\jd{on-chip} and off-chip) voltage regulation.
2) {\tech} has a \ja{very small} \jkv{(i.e., less than $1\%$)} performance degradation compared to LDO and MBVR PDNs \jc{(the highest performing PDNs \jkc{at} $4W$ TDP)}. \jkz{{\tech} performs \jkv{only} slightly worse than the LDO \lo{and MBVR PDNs} \lo{due to} {\tech}'s higher load-line that \lo{is a result of} resource sharing \lo{between} \jkc{its LDO and IVR components} \lo{within {\tech}'s} hybrid PDN} (discussed in Sec. \mbox{\ref{sec:new_pdn}}). 
\jc{3) The I+MBVR \jkz{PDN} provides higher performance than \jkz{the IVR PDN} \lo{($6\%$ on average)} since I+MBVR \jkz{removes the two-stage} voltage regulation of \jkz{the} SA and IO domains. \jkz{This change improves the ETEE of the I+MBVR PDN over the IVR PDN, and therefore increases} the power-budget of \jkz{the} CPU core domain.}
\jd{4) The performance improvement of the five PDNs correlates with the performance-scalability of the workloads}\jkc{, since the performance-scalability metric reflects how the performance of an application improves as the CPU clock frequency increases (due to the additional power-budget allocated to the CPU cores).} 

\lo{We} conclude that {\tech} significantly improves the CPU core performance compared to the state-of-the-art PDN (IVR) at a low TDP point  by operating in \lo{\texttt{LDO-Mode},} which results in a higher ETEE than that of the IVR PDN}.

\begin{figure*}[h]
 \vspace*{-5mm}
  \begin{center}
  \includegraphics[trim=0.85cm .8cm .95cm 0.8cm, clip=true,width=1\textwidth,keepaspectratio]{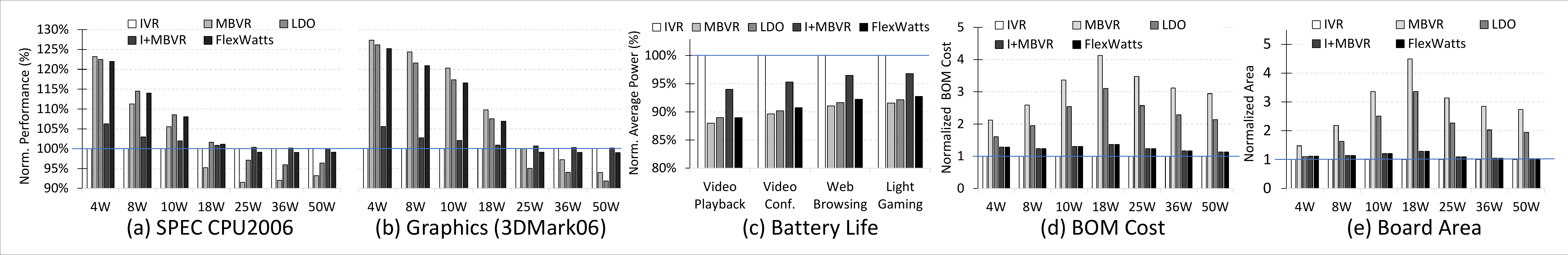}\\
  \caption{Evaluation of the five PDNs normalized to IVR PDN (the state-of-the-art PDN~\cite{2_burton2014fivr,5_nalamalpu2015broadwell,icelake2020}) (a) SPEC \jc{CPU2006} average performance, (b) 3DMark06 performance, (c) Battery life workloads, (d) BOM, and (e) Board \jkc{area}.}\label{fig:all_flexwatts_res}    
  \end{center} 
 \vspace*{-4mm}
\end{figure*}

\noindent \textbf{SPEC \jc{CPU2006} \jkz{Benchmarks} at 4W to 50W TDP.}
We examine the effects of using different processor \jc{TDP} levels, ranging from $4W$ to $50W$\jc{, on CPU performance.} \jc{Fig.~\ref{fig:all_flexwatts_res}(a) plots \jkc{the average performance across the SPEC CPU2006 benchmarks for several TDP levels}}. 
Based on Fig. \ref{fig:all_flexwatts_res}(a), we make \jc{three} key observations. 
\jkc{1) At TDPs \emph{lower} than $18W$, {\tech} provides up to $22\%$ higher performance over the IVR PDN by operating mainly in \texttt{LDO-Mode}, which has a \emph{higher} ETEE than the IVR PDN at \emph{low} TDPs. Compared to the highest-performing PDNs (MBVR/LDO) at \emph{low TDPs}, {\tech} performs \jkv{only slightly  (i.e., less than $1\%$)} worse due to the \emph{higher} load-line of {\tech}'s \texttt{LDO-Mode}. 
2) At TDPs \emph{higher} than $18W$, {\tech} provides up to $7\%$/$4\%$ higher performance over the MBVR/LDO PDNs by operating mainly in \texttt{IVR-Mode}, which has a \emph{higher} ETEE than the MBVR/LDO PDNs at \emph{high} TDPs. Compared to the highest-performing PDN (IVR) at \emph{high TDPs}, {\tech} performs \jkv{only slightly (i.e., less than $1\%$)} worse  due to the \emph{higher} load-line of {\tech}'s \texttt{IVR-Mode}}.
\jc{3) The I+MBVR PDN provides higher (up to $6\%$) performance than \jkz{the IVR PDN} across the \jkz{tested} TDP range since I+MBVR \jkz{removes the two-stage} voltage regulation of \jkz{the} SA and IO domains. \jkz{This change improves the ETEE of the I+MBVR over the IVR PDN, and therefore increases} the power-budget of \jkc{the} CPU core domain. 
However, I+MBVR provides significantly lower performance (up to $15\%$) than {\tech} \lo{at low TDPs,} since the I+MBVR PDN uses \jkz{two-stage} voltage regulation (\je{i.e., for the CPU
cores, LLC, and graphics domains}), \jkz{which results in a lower ETEE compared to {\tech}} at low TDPs (e.g., $4W$)}.

\noindent \textbf{Graphics Workloads at 4W to 50W TDP.}
We evaluate different PDN architectures using the 3DMark06 graphics \jd{workloads}~\cite{17_3DMARK}. \jkc{While running these workloads, $10\%$ to $20\%$ of the processor's power-budget is allocated to the \jc{CPU} cores, while the rest is allocated to the graphics engines.} In addition, since the \jkc{graphics workloads \jkv{require high} memory bandwidth, the LLC domain operates} at a higher frequency and \jkv{higher} voltage than the \jkv{CPU domain}.

Fig. \ref{fig:all_flexwatts_res}(b) shows the \jd{average} performance of the 3DMark06 graphics \jd{workloads} with the \jc{five} PDN architectures when running at $4W$ to $50W$ TDP. We make \lo{four} key observations. 
\jkc{1) At TDPs \emph{lower} than $25W$, {\tech} provides up to $25\%$ higher performance over the IVR PDN by operating mainly in \texttt{LDO-Mode}, which has a \emph{higher} ETEE than the IVR PDN at \emph{low} TDPs. 
\lo{2)} At TDPs \emph{higher} than $25W$, {\tech} provides up to $3\%$/$6\%$ higher performance over MBVR/LDO PDNs by mainly operating in \texttt{IVR-Mode}, which has a \emph{higher} ETEE than the MBVR/LDO PDNs at \emph{high} TDPs.
3) \jc{{\tech} performs slightly worse (i.e., up to $2\%$ lower) than \lo{MBVR/LDO} PDNs due to i) the higher load-line of {\tech}, and \lo{ii)} the large difference in operating voltages across the CPU \lo{core}, LLC and \lo{graphics} domains while running graphics workloads (i.e., \lo{the core domain requires} low voltage, e.g., $0.5V$, while graphics domain requires high voltage, e.g., $0.9V$), which degrades the ETEE of both {\tech} (in \texttt{LDO-Mode}) and LDO PDNs (as we discuss in Sec. \ref{pdn_bg}).}}
\hjb{\lo{4)} The I+MBVR PDN provides \lo{up to $6\%$} higher \jkc{performance than the IVR PDN across the tested} TDP range. \lo{I+MBVR} improves the power conversion efficiency for the SA and IO domains \jkc{(which results in I+MBVR having a higher ETEE than the IVR PDN), and increases} the power-budget of \lo{the graphics} domain. However, I+MBVR provides significantly lower performance (up to $19\%$) than {\tech} \jkc{at \emph{low} TDPs, since the I+MBVR PDN's two-stage voltage regulation (similar to IVR PDN) at low TDPs (e.g., $4W$) results in \jkc{a lower ETEE than}}} {\tech}.

\hjp{Based on our extensive \ja{CPU-} and graphics\ja{-intensive} workload \jkz{evaluations}, we \textbf{conclude} that {\tech} increases \jkc{the performance \jd{of a low TDP (e.g., $4W$) processor} by \jd{up to $25\%$,} while maintaining a low (i.e., less than $2\%$) performance degradation \jd{for high TDP processors compared} to the state-of-the-art IVR PDN, over a wide range of TDPs (i.e., \je{$4W$--$50W$})\jkv{. This is because} {\tech} \ja{1) allocates} the hybrid PDN to domains with a wide power consumption range \jkc{(i.e., CPU cores, LLC, and graphics)}, thereby maintaining a high ETEE across the wide \jd{power} range, and 2) \jc{allocates} an off-chip VR to each domain with \jkc{a} \jb{low and} narrow power consumption range (i.e., SA and IO), thereby maintaining high power conversion efficiency \jc{in} these \lo{domains,} which \jkc{increases {\tech}'s} ETEE across \jkv{\emph{all}} TDPs and workloads compared to \jkc{the IVR PDN}}.} 
\noindent \textbf{Battery Life Workloads.}
We choose four workloads that are \jc{commonly} used to evaluate the battery life  of mobile processors~\lo{\cite{zhang2017race,19_MSFT,boroumand2018google}:  video playback~\cite{boroumand2018google,19_MSFT}}, video conferencing~\cite{mobilemark,boroumand2018google}, web browsing~\cite{sysmark,mobilemark}, and light gaming  \cite{shaker2013evolving} benchmarks. For our modeled system, \lowercase{video playback, video conferencing, web browsing, and light gaming} have $10\%$, $20\%$, $30\%$, and $40\%$ active state \jc{with minimum frequency} ($C0_{MIN}$) residencies, respectively. During \lo{the} remaining execution time, compute domains (cores, LLC, and graphics engines) are idle\jkc{,} \lo{but} the system agent \ja{(SA)} has activity at the display-controller \ja{(in package-C8 state)} and \jc{performs} periodic \ja{(every few hundreds of microseconds)} memory accesses \ja{(in package-C2 state)}. We note that these workloads have nearly the same average power consumption regardless \jc{of} the TDP of \lo{the} system. In active and idle states, we assume the same nominal power at all TDPs. We evaluate battery life workloads at \ja{$T_j$} of $50^{\circ}C$.
Fig. \ref{fig:all_flexwatts_res}(c) shows the average (normalized to IVR) power consumption of \jc{the} five PDNs. \hjp{We observe that {\tech} consumes up to $1\%$ more power than MBVR, but $8\%$ to $11\%$ less \lo{power} than IVR when running the four battery life workloads.} \hjb{I+MBVR \lo{consumes} up to $6\%$ \lo{less} average power than IVR and $5\%$ higher average power than {\tech}.}

We conclude that {\tech} is almost as energy-efficient as both MBVR and LDO and up to $11\%$ more energy-efficient than IVR, for battery life workloads. \jc{This is mainly because, in low power states (i.e., package C-states) and \lo{the low-frequency} active state (i.e., $C0_{MIN}$) of the battery life workloads, {\tech} operates \lo{in} \lo{\texttt{LDO-Mode},} which has better power conversion efficiency that \lo{IVR} in these low power consumption states, thereby maintaining high power conversion efficiency across battery life workloads.}

\noindent \textbf{\lo{BOM.}}
Fig. \ref{fig:all_flexwatts_res}(d) shows the BOM of the five PDNs normalized to IVR for $4W$--$50W$ TDPs. \hjp{\lo{We} make two key observations. 1) {\tech} and} \hjb{I+MBVR} PDNs have comparable cost to IVR. 2) MBVR and LDO have $2.1\times$--$4.2\times$ and $1.6\times$--\je{$3.1\times$} higher BOM, respectively, compared to \jkv{IVR,} across the wide TDP range.  


\noindent \textbf{\lo{Board Area.}} Fig. \ref{fig:all_flexwatts_res}(e) shows board area of the five PDNs normalized to IVR for the $4W$--$50W$ TDP range. We make two key observations. 1) {\tech} and \hjb{I+MBVR}\hjp{ have comparable board area to IVR. 2) MBVR and LDO have $1.5\times$--$4.5\times$  and $1.1\times$--$3.3\times$ higher area, respectively, \lo{compared} to IVR.}

\noindent \textbf{Why does {\tech} have better BOM and board area than LDO and MBVR?} The advantage of {\tech} in BOM and board area over MBVR and LDO is due its \emph{reduced maximum-current}, $Icc_{max}$. This happens due to two reasons. First, {\tech} uses a shared voltage regulator for the high power domains (i.e., cores, graphics, and LLC), which enables current sharing between these three domains. Second, {\tech} has reduced current (by nearly 50\%) \lo{in} \texttt{IVR-Mode} compared to LDO, and as such, the shared VR (between the cores, graphics, and LLC) is designed with a maximum-current level similar to \lo{that of} IVR. When a high power (and \lo{thus high} current) workload (e.g., Turbo Boost \cite{rotem2015intel}) is requested, the hybrid PDN \emph{switches} to \lo{the} \texttt{IVR-Mode}, and thus {\tech} has comparable maximum-current to IVR.

We conclude that \ja{{\tech} provides significant performance and energy improvements with a \jkv{low} BOM and area overhead compared to the state-of-the-art PDN, over a wide power consumption range and a wide variety of workloads.}

\section{Related Work}\label{sec:related}

\jkc{To our knowledge, this is the first work to 1) provide a versatile framework, {\modl}, that enables multi-dimensional architecture-level exploration of modern processor \jkv{power delivery networks (PDNs)}, and 2) propose a novel adaptive hybrid PDN, {\tech}, that \jkv{provides} high efficiency and performance in client processors across a wide spectrum of power consumption and \jkv{workloads, compared to four state-of-the-art PDNs~\cite{5_nalamalpu2015broadwell,tam2018skylake,burd2019zeppelin,skylakex}, as we demonstrate both qualitatively and quantitatively}. We discuss other related works here.}

\jc{A recent work~\cite{li2018workload} proposes an adaptive PDN  that can dynamically manage on/off-chip VRs in hybrid PDN systems based on the \jkx{dynamic} workload. The proposed solution uses many \lon{on-chip} and off-chip VRs, and targets many-core systems  that are optimized for only a single TDP.  \lon{Unlike {\tech}}, this solution is not optimized for cost, area, or client \jkx{(laptop and desktop) systems}.

Many \lon{existing} works \lon{investigate} the potential of integrated VRs~\lon{\cite{2_burton2014fivr,6_kanter2013haswell,16_varma2015power,23_wang2015analytical,24_li2017energy,yan2012agileregulator}}. \lon{PowerSoC~\cite{23_wang2015analytical} is an analytical model of a PDN system that includes on-chip VRs, off-chip VRs, and PDN models}, providing a platform to evaluate \lon{performance} and explore the design space of the entire PDN system. \lon{The authors} show that hybrid PDN architectures with both on-chip and off-chip VRs can achieve a better tradeoff between area and efficiency requirements compared to traditional off-chip paradigms.  Haoran \textit{et al.} \cite{24_li2017energy} \jkx{compare} the characteristics of different PDNs for many-core systems using on-chip and/or off-chip VRs \lon{using} an analytical model. 
Yan \textit{et al.} \cite{yan2012agileregulator} \jkx{propose} a hybrid PDN that \jc{optimizes the} area-energy tradeoff to improve the energy-efficiency of multi-core architectures by using several redundant cores powered by dedicated on-chip or off-chip VRs and migrating workloads that can benefit from fast DVFS to cores powered by on-chip VRs. \lon{Other works \cite{2_burton2014fivr,6_kanter2013haswell,16_varma2015power} claim that the fully-integrated voltage \lon{regulator,}  first adopted in Intel's~4th generation Core \lon{processors}~\cite{2_burton2014fivr}, improves \lon{performance} and increases battery life in client systems.}
\jkx{These works} have at least one of two main \jkx{shortcomings}.
First, several of these prior works \cite{23_wang2015analytical,yan2012agileregulator,24_li2017energy} are not optimized for \jkx{three key design parameters for client processors:} cost, area, or different TDPs. Second, some works  \cite{2_burton2014fivr,6_kanter2013haswell,16_varma2015power} do not \jd{address the inefficiencies of the IVR PDN  in terms of performance (e.g., \jkc{at} \jkv{low TDPs}) and energy (e.g., for \jkc{computationally-light} workloads), which makes these works inefficient for client processors across a wide power and workload range.}}
\jk{Compared to all aforementioned \lon{works}, our \lon{experimental} \lon{study} 1) models a wide TDP range, showing which PDN is better for high performance and high energy efficiency at each TDP \lon{level, and} 2) evaluates \lon{a wide variety of} mobile client system workloads, \lon{providing} an understanding of which PDN architecture is more efficient for each workload.}  

\section{Conclusion}\label{sec:summary}
In this work, we \jkz{first} develop {\modl}, a framework that enables architectural exploration of \jkz{power delivery network (PDN)} architectures with respect to multiple metrics: performance, battery life, BOM and board area.  
\jk{Using {\modl}, we observe multiple energy inefficiencies in the PDNs of recent  client processors. We introduce a new power- and workload-aware hybrid PDN, {\tech}, to improve \jkx{the} performance and energy-efficiency of  client processors for a wide \jkz{power and workload} range.} 
\jk{We provide a practical implementation of {\tech}, where we design a mode-switching power management flow that guarantees to switch the \jkx{hybrid PDN safely} between two PDN modes, without \jkx{undesirable} voltage noise. We  present a new algorithm that automatically switches {\tech} to the PDN mode that results in the \jkz{highest} energy-efficiency, battery life, and performance. Our evaluations show that {\tech} provides significant performance and energy improvements with a \ja{very small} BOM and area overhead compared to the state-of-the-art PDN, over a wide power consumption range and a wide variety of workloads. 
\ja{We hope that our open-source release of {\modl} fills a gap in the space of publicly-available experimental PDN infrastructures and\jkz{, along with {\tech},} inspires new studies, ideas, and methodologies in PDN system design.}}


